%% file: main_clean.tex
\documentclass[a4paper,fleqn,usenatbib,useAMS]{mnras}
\usepackage{graphicx,amssymb,amsmath,anysize,mathrsfs}
\usepackage[T1]{fontenc}
\usepackage{aecompl}
\usepackage{float}
\usepackage{multirow}
\usepackage{booktabs}
\usepackage{appendix}
\usepackage{enumerate}
\usepackage{comment}
\usepackage{dsfont}
\usepackage{braket}
\usepackage[utf8]{inputenc}
\usepackage{color}
\usepackage[usenames,dvipsnames]{xcolor}
\usepackage{fancyhdr}
\usepackage{mathtools}
\usepackage{lscape}
\usepackage{caption}
\usepackage{subcaption}
\usepackage[font=small,labelfont=bf]{caption}
\usepackage{slashed}
\usepackage{arydshln}
\usepackage{listings}
\usepackage{wrapfig}
\usepackage{soul}

\usepackage{pbox}
\usepackage[pagewise]{lineno}
\newcommand{\fom}{\text{FoM}_{\omega}}
\newcommand{\fomg}{\text{FoM}_\gamma}
\newcommand{\fomc}{\text{FoM}_{\gamma\omega}}
\newcommand{\fomdetf}{\text{FoM}_{\text{DETF}}}

\newcommand{\xfigure}[3]{
\begin{figure}
\centering
\hspace*{-4.5mm}   
\includegraphics[width=0.50\textwidth]{figure/#1}
\caption{#2}
\label{#3}
\end{figure}
}

\newcommand{\xfigurefull}[4]{
\begin{figure*}
\centering
\includegraphics[width=#1\textwidth]{figure/#2}
\caption{#3}
\label{#4}
\end{figure*}
}

\newcommand{\xxfigure}[4]{
\begin{figure}
\centering
\hspace*{-4.5mm}   
	\begin{subfigure}[b]{0.52\textwidth}
	\includegraphics[width=1\textwidth]{figure/#1}
	\end{subfigure}
	\hspace*{-4.5mm}   
	\begin{subfigure}[b]{0.52\textwidth}
	\includegraphics[width=1\textwidth]{figure/#2}
	\end{subfigure}
\caption{#3}
\label{#4}
\end{figure}
}

\begin{document}

\title{Cosmological constraints from multiple tracers in spectroscopic surveys}

\author[A. Alarcon, M. Eriksen, E. Gaztanaga]{
Alex Alarcon$^{1}$\thanks{E-mail: \href{mailto:alarcon@ice.cat}{alarcon@ice.cat (AA);}},
Martin Eriksen$^{2}$\thanks{E-mail: \href{mailto:marberi@strw.leidenuniv.nl}{marberi@strw.leidenuniv.nl (ME);}},
Enrique Gaztanaga$^{1}$\thanks{E-mail: \href{mailto:gaztanaga@gmail.com}{gaztanaga@gmail.com (EG)}}\\ \\
$^{1}$Institut de Ciències de l’Espai (IEEC-CSIC), E-08193 Bellaterra (Barcelona), Spain\\
$^{2}$Leiden Observatory, Leiden University, PO Box 9513, NL-2300 RA Leiden, Netherlands}

\maketitle

\input{abstract_clean}


\input{intro_clean}

\input{modeling_clean}

\input{samplevar_clean}

\input{hod_clean}

\input{partoverlap_clean}

\input{conclusions_clean}

\input{acknowledgements_clean}

\input{appendix_A_clean}

\input{appendix_B_clean}

\bibliography{biblio}{}
\bibliographystyle{mn2e}

\end{document}

%% file: abstract_clean.tex
\begin{abstract}

We use the Fisher matrix formalism to study the expansion and growth history of the Universe using galaxy clustering with 2D angular cross-correlation tomography in spectroscopic or high resolution photometric redshift surveys.
The radial information is contained in the cross correlations between narrow redshift bins. 
We show how multiple tracers with redshift space distortions cancel sample variance and arbitrarily improve the constraints on the dark energy equation of state $\omega(z)$ and the growth parameter $\gamma$ in the noiseless limit. 
The improvement for multiple tracers quickly increases with the bias difference between the tracers, up to a factor $\sim4$ in $\fomc$. 
We model a magnitude limited survey with realistic density and bias using a conditional luminosity function, finding a factor 1.3-9.0 improvement in $\fomc$ -- depending on global density -- with a split in a halo mass proxy. 
Partly overlapping redshift bins improve the constraints in multiple tracer surveys a factor $\sim1.3$ in $\fomc$. This findings also apply to photometric surveys, where the effect of using multiple tracers is magnified. We also show large improvement on the FoM with increasing density, which could be used as a trade-off to compensate some possible loss with radial resolution.

\end{abstract}

%% file: intro_clean.tex
\section{Introduction} \label{sec:introduction} 
\setlength{\parindent}{6ex}
One of the most exciting and enigmatic discoveries in the recent years is the late time accelerated expansion of the Universe, confirmed in late 1990s from Type Ia supernovae (\citealt{1998AJ....116.1009R}, \citealt{1999ApJ...517..565P}). During the last decade a wide range of observations (see \citealt{2013PhR...530...87W}) has provided robust evidence for cosmic acceleration, consistent with a $\Lambda$CDM model dominated by a new component called \textit{dark energy}, which properties and origin remain unknown. 

Cosmic expansion is parametrized by $\Omega(a)$ and the DE equation of state $\omega(a)=\omega_0 + \omega_a (1-a)$ (\cite{2001IJMPD..10..213C}, \cite{2003PhRvL..90i1301L}) while cosmic growth is parametrized by $\gamma$, which gives the growth rate as $f(z)= \Omega(z)^\gamma$. For General Relativity (GR) $\gamma\sim0.55$, while Modified Gravity models can give different values of  $\gamma$ for the same expansion history (e.g. \citealt{2001ApJ...548...47G}, \citealt{2004PhRvD..69d4005L}, \citealt{2015APh....63...23H}).  Here we study the dark energy equation of state $\omega(z)$ and growth rate $\gamma$ constraints using galaxy clustering in spectroscopic surveys. Galaxy clustering is able to probe the expansion and growth history almost independently, unlike weak lensing surveys alone, which are limited to projected, 2D information (see \citealt{2012MNRAS.422.2904G}, \citealt{2013PhR...530...87W}). Galaxies are easy to observe and by accurately measuring their redshift one can reconstruct the 3D clustering information.

Unfortunately, the relation between galaxy and dark matter is not straight-forward, and in the linear regime, for large scales, it can be modeled by a factor called \textit{linear bias} $b(k, z)$, such that $\delta_g(k,z)=b(k, z) \delta_m(k,z)$, where $\delta_g$ and $\delta_m$ are galaxy and dark matter fluctuations. An independent measurement is needed to break the degeneracy between bias and $\gamma$, as galaxy clustering alone cannot (e.g. see Eq.\ref{xithetaz} below). One can break this degeneracy using cross-correlation with lensing surveys (e.g. \citealt{2012MNRAS.422.2904G}, \citealt{2013PhR...530...87W}), but in this paper we will focus on spectroscopic surveys or high resolution photometric surveys (\citealt{2014MNRAS.442...92M}). In this case, to determine bias one can measure the redshift space distortion parameter $\beta\equiv f(z)/b(z)$. Redshift space distortions (RSD) in the linear regime (\citealt{1987MNRAS.227....1K}) enhance clustering in the line of sight by a factor $(1+f)$ due to local infall of bodies as a result of gravity. Measuring with different angles relative to the line of sight one can determine $f(z)$. However, the random nature of fluctuations (sampling variance) limits the accuracy with which one can determine $P(k)$, and with only one tracer that propagates to $\beta$ and cosmological parameters. \cite{2008.0810.0323} proposed to use multiple tracers of the same underlying distribution to beat this limit measuring along many directions and improve the constraints canceling sampling variance with RSD. Sampling variance cancellation can also be achieved with other observables (e.g. \citealt{2004MNRAS.350.1445P}, \citealt{2009PhRvL.102b1302S}).
This technique has been explored in recent literature (e.g. \citealt{2009MNRAS.397.1348W}, \citealt{2010MNRAS.407..772G}, \citealt{2011MNRAS.416.3009B}, \citealt{2012MNRAS.420.2042A}, \citealt{2016MNRAS.455.3871A}),  also for photometric surveys (\citealt{2014MNRAS.445.2825A}) and combining lensing and spectroscopic surveys (\citealt{2012MNRAS.422.1045C}, \citealt{2015MNRAS.451.1553E}).

We use 2D angular correlations $C_\ell$ (see $\S$\ref{subsec:corr}) to avoid assuming a cosmology and avoid overcounting overlapping modes without including the full covariance between them (\citealt{2014arXiv1412.2208E}). We forecast spectroscopic surveys with narrow redshift bins ($\Delta z=0.01(1+z)$) such that the radial linear modes will be in the cross correlations between redshift bins. In the fiducial forecast we will compute the correlations using redshift space distortions (RSD) and we include baryon acoustic oscillation measurements (BAO). In this paper we will study the constraints from single spectroscopic tracers as compared to splitting one population into two tracers. The single tracers are denoted as B1 and B2 and the multiple tracer survey as B1xB2. Note this differs from our series of previous studies (\citealt{2014arXiv1412.2208E} and others) where we included one Bright and one Faint population as opposed to two Bright populations. The cosmological parameter error estimation is done using the Fisher matrix formalism described in $\S$\ref{subsec:fisher}, and we quantify the relative strength of the surveys through the Figures of Merit (FoMs) defined in $\S$\ref{subsec:fom}, which focus on measuring the expansion and growth history simultaneously. In subsection \ref{subsec:fiducial} we present our fiducial forecast assumptions.

This paper is organized as follows. In section 2 we present our modeling and fiducial forecast assumptions. Section 3 discusses sample variance cancellation in surveys with multiple tracers and explores the effect of the relative bias amplitude between two tracers and the dependence on galaxy density. In section 4 we model galaxy bias using a conditional luminosity function (CLF) and halo model to build an apparent limited survey to study the tradeoff between galaxy bias and galaxy density when we split a survey into two subsamples. Section 5 investigates the impact of having partly overlapping redshift bins between two tracers in a multi tracer survey and how this affects the constraints. Moreover, it studies radial resolution by increasing the number of redshift bins. In section 6 we present our conclusions. Appendix A studies the importance of RSD and BAO in the constraints and the degeneracy with cosmological parameters. Appendix \ref{appB} shows the dependence that the constraints have on the bias evolution in redshift. 

In this paper we have produced the results with the forecast framework developed for \cite{2012MNRAS.422.2904G}, \cite{2014arXiv1412.2208E}, \cite{2015MNRAS.451.1553E}, \cite{2015arXiv150203972E} and  \cite{2015arXiv150800035E}.

%% file: modeling_clean.tex
\setlength{\parindent}{0ex}

\section{Modeling and forecast assumptions} \label{sec:theobkg}  

\subsection{Galaxy bias} \label{subsec:bias}

In the local bias model (\citealt{1993ApJ...413..447F}), where fluctuations are small, one can approximate the relation between galaxy overdensities $\delta_g$ to matter overdensities $\delta_m$ through

\begin{equation} \label{galaxybias}
\delta_g(k,z) = b(k,z) \delta_m (k,z)
\end{equation} 

\medskip

where $b(z,k)$ is the galaxy bias, which can in general depend on the scale and redshift. It also varies between different galaxy populations (galaxies hosted by more massive haloes tend to be more biased, eg. \citealt{2001ApJ...546...20S}). Then, for scale independent bias $b(z) = b(k,z)$ the angular correlations $\xi_{gg} \equiv \langle\delta_g \delta_g\rangle$ we have that

\begin{equation} \label{xithetaz}
\xi_{gg}(\theta, z) = b^2(z)\, \xi_{mm}(\theta, z) \propto b^2(z)\ D^2(z)
\end{equation}

\medskip

Galaxy bias can also include an stochastic component $r$, see also Eq. \ref{b1}, which is also a common measure of non-linearity

\begin{equation}
r \equiv \frac{\xi_{gm}}{\sqrt{\xi_{gg}\xi_{mm}}}.
\end{equation} 

\medskip

In \cite{2012MNRAS.422.2904G} it was shown that it can be treated as a re-normalisation of bias in large scales and here it is fixed to $r=1$. In addition, non local bias can also modify the galaxy correlation function, but this is a smaller effect (\citealt{2012PhRvD..85h3509C}).
\setlength{\parindent}{6ex}

We include \textit{redshift space distortions} (RSD, redshift displacement of galaxies due to their peculiar velocities with respect to the comoving expansion) using linear theory, \cite{1987MNRAS.227....1K}, assuming no velocity bias

\begin{equation} \label{rsdfourierbias}
\delta_s(k,\mu) = (b+f\mu^2)\,\delta(k).
\end{equation} 
\setlength{\parindent}{0ex}

where $\mu\equiv(\hat{z}\cdot\boldsymbol{k})/k=k_\|/k$. We define $\beta\equiv f/b$ as the term with specific angular dependence $\mu$ in redshift space.

\subsection{Angular correlation function and Power spectrum} \label{subsec:corr}

Consider the projection of spatial galaxy fluctuations $\delta^{i}_g(x,z)$ along a given direction in the sky $\hat{\boldsymbol{r}}$
\begin{equation}
\delta^{i}_g(\hat{\boldsymbol{r}}) = \int dz \, \phi^{i}(z) \, \delta^{i}_g(\hat{\boldsymbol{r}},r, z),
\end{equation} 

\medskip

where $\phi^{i}(z)$ is the radial selection function for the i-th redshift bin of a given tracer. We define the angular correlation between galaxy density fluctuations as 
\begin{equation}
\omega_{ij}(\theta)\equiv\langle\delta^{i}_{g}(\boldsymbol{r}) \,\delta^{j}_{g}(\boldsymbol{r}+\hat{\theta})\rangle.
\end{equation}

\medskip

Expanding the projected density in terms of spherical harmonics we have

\begin{equation}
\delta^{i}(\hat{\boldsymbol{r}})=\sum_{\ell\geq0} \sum_{m=-\ell}^{\ell} a^{i}_{\ell m}Y_{\ell m}(\hat{\boldsymbol{r}})
\end{equation} 

\medskip

where $Y_{\ell m}$ are the spherical harmonics. The coefficients $a^{i}_{\ell m}$ have zero mean $\braket{ a^{i}_{\ell m}}=0$, as $\braket{\delta^{i}}=0$ by construction, and their variance form the angular power spectrum
\begin{equation} \label{ellvariance}
\braket{ a^{i}_{\ell m}  a^{j}_{\ell' m'}} \equiv \delta_{\ell\ell'}\delta_{mm'} \, C^{ij}_{\ell}
\end{equation} 

which can be related to the angular correlations with

\begin{equation}
\omega_{ij}(\theta) = \sum_{\ell\geq0} \frac{2\ell+1}{4\pi} \,L_\ell(\cos{\theta}) \, C^{ij}_\ell
\end{equation} 

\medskip

where $L_\ell(\cos{\theta})$ are the Legendre polynomials of order $\ell$. The $C^{ij}_\ell$ can be expressed in Fourier space (\citealt{2011MNRAS.414..329C}) as 

\begin{equation} \label{angcorr}
C_\ell^{ij} = \frac{1}{2\pi^2} \int 4\pi k^2 dk \, P(k)\, \psi^i_\ell (k) \psi^j_\ell (k)
\end{equation} 

\medskip

where $P(k)$ is the matter power spectrum and $\psi^i_\ell (k)$ is the kernel for the i-th redshift bin of a given population. For the matter power spectrum $P(k)$ we use the linear power spectrum from \citealt{1998ApJ...496..605E} for linear scales, which accounts for baryon acoustic oscillations (BAO). In real space (no redshift space distortions), taking into account only the intrinsic component of galaxy number counts, this kernel is (\citealt{2014arXiv1412.2208E})

\begin{equation}
\psi^{i}_\ell(k) = \int dz \, \phi^{i}(z)\, D(z)\, b(z,k)\, j_\ell(kr(z))
\end{equation} 

\medskip

where $b(z,k)$ is the galaxy bias, Eq. $\ref{galaxybias}$. When including RSD, one has to add an extra term that in linear theory is given by (\citealt{1987MNRAS.227....1K}, \citealt{1994MNRAS.266..219F}, \citealt{1995ApJS..100...69F}, \citealt{1995AIPC..336..381T})

\begin{equation}
\begin{split}
\psi_\ell(k) =& \psi_\ell^{Real} + \psi_\ell^{RSD} \\
\psi_\ell^{RSD} =& \int dz\, f(z) \, \phi(z)\, D(z)\,\left[ L_0(\ell) \,j_\ell(kr)\right.\\
&\left.+L_1(\ell) \,j_{\ell-2}(kr) + L_2(\ell) \,j_{\ell+2}(kr)\right]
\end{split}
\end{equation} 

\medskip

where $f(z)$ is the growth rate and 
\begin{equation}
\begin{split}
L_0(\ell) &\equiv \frac{(2\ell^2+2\ell-1)}{(2\ell+3)(2\ell-1)}\\
L_1(\ell) &\equiv -\frac{\ell(\ell-1)}{(2\ell-1)(2\ell+1)} \\
L_2(\ell) &\equiv -\frac{(\ell+1)(\ell+2)}{(2\ell+1)(2\ell+3)}
\end{split}
\end{equation} 

The fiducial modeling includes RSD in the kernel and BAO in the power spectrum, but we will also forecast removing one or both of these effects.

\subsubsection{Covariance}

Angular cross correlations between a redshift bin $i$ and redshift bin $j$ correspond to the variance of spherical harmonic coefficients $a_{\ell m}$ (Eq. \ref{ellvariance}). Assuming that $a^{i}_{\ell m}$ are Gaussianly distributed and in a full sky situation, one can then estimate each $\ell$ angular power spectrum using the $2\ell+1$ available modes,
\begin{equation}
\tilde{C}^{ij}_{\ell} = \frac{1}{2\ell+1}\sum_{m=-\ell}^{\ell} a^{i}_{\ell m}  a^{j}_{\ell m}.
\end{equation}

\medskip

which yields Eq. \ref{angcorr}. However, in a more realistic situation, we only have partial coverage of the sky so that the different modes $\ell$ become correlated. Following the approach of \cite{2007MNRAS.381.1347C}, we bin the $\ell\times\ell$ covariance with a sufficiently large band width $\Delta \ell$ such that it becomes  band diagonal, and scale it with $1/f_{Sky}$ (where $f_{Sky}$ is the survey fractional sky). Then, the covariance becomes

\begin{equation}
\text{Cov}\,[ \hat{C}^{ij}_{\ell} , \hat{C}^{kl}_{\ell}] = N^{-1}(\ell) (\hat{C}^{ik}_{\ell} \hat{C}^{jl}_{\ell} + \hat{C}^{il}_{\ell} \hat{C}^{jk}_{\ell}) .
\end{equation} 

\medskip 

where $N(\ell)=f_{Sky}(2\ell+1)\Delta \ell$, and the correlation $\hat{C}$ includes observational noise 

\begin{equation} \label{shotnoise}
\hat{C}^{ij}_{\ell} = C^{ij}_{\ell} + \delta_{ij}\, \frac{1}{\bar{n}_g}
\end{equation} 

\medskip

where $\bar{n}_g=\frac{N_g}{\Delta\Omega}$ is the galaxy density per solid angle. The first term in Eq. \ref{shotnoise} is  signal and contains sample variance information, while the second is shot-noise.  Then, we can define the $\chi^2$ as

\begin{equation} \label{chi}
\begin{split}
\chi^2 = \sum_{\ell,\ell',i,j}\left(C^{ij}_{\ell}(\{\lambda_k\}) - \hat{C}^{ij}_{\ell}\right) &\left(\text{Cov}^{-1}\right)_{\ell,\ell',i,j} \\
&\times\left(C^{ij}_{\ell'}(\{\lambda_k\}) - \hat{C}^{ij}_{\ell'}\right)
\end{split}
\end{equation} 

\medskip

where $C^{ij}_\ell(\{\lambda_k\})$ depend on the parameters that we are looking for, $\hat{C}^{ij}_{\ell}$ are the observed $C^{ij}_\ell$'s and Cov the covariance matrix between the $\hat{C}^{ij}_\ell$'s. $C^{ij}_\ell$ include both auto and cross correlations between different redshift bins and also possibly between different tracers if there was more than one.

\subsubsection{Nonlinearities}

As we are working in the linear regime we have to limit the scales that we include in the forecast. We restrict the forecast to scales between $10\leq \ell \leq 300$. In addition we apply a further cut in $l_{max}$ 

\begin{equation} \label{kmax}
\ell_{max} = k_{max}\, r(z_i) -0.5,
\end{equation} 

for which correlations to include, as these are the scales contributing to $C(\ell)$ for a given narrow redshift bin $z_i$ (\cite{2014arXiv1412.2208E}). In the forecast we use the Eisenstein-Hu power spectrum and the MICE cosmology with a maximum scale $k_{max}$ of (see \citealt{2015arXiv150203972E})
\begin{equation} 
k_{max}(z) = \exp{(-2.29 + 0.88z)}.
\end{equation}

\subsection{Fisher matrix formalism}\label{subsec:fisher}

Even if we don't have any data, we can tell how $\chi^2(\{\lambda_\mu\})$ will vary in the parameters space defined by $\{\lambda_\mu\}$. Expanding $\chi^2$ in the Gaussian approximation around its minimum the Fisher matrix is (\citealt{1935...Fisher}, \citealt{Dodelson})

\begin{equation}
F_{\mu\nu} =  \sum_{l,l'} \sum_{ij,mn}\frac{\partial C^{ij}_{l}}{\partial\lambda_\mu} \left(\text{Cov}^{-1}\right)_{l,l'} \frac{\partial C^{mn}_{l'}}{\partial\lambda_\nu},
\end{equation} 

\medskip

and it follows that
\begin{equation} \label{lambdacov}
\text{Cov}\left[\lambda_\mu, \lambda_\nu\right] = \left[ F^{-1}\right]_{\mu\nu}.
\end{equation} 

\medskip

The parameters included in the Fisher matrix forecast are (\citealt{2015arXiv150203972E})
\begin{equation}
\{\lambda_\mu\} = \omega_0, \omega_a, h, n_s, \Omega_m, \Omega_b, \Omega_{DE}, \sigma_8, \gamma, \text{Galaxy bias}.
\end{equation} 

The forecast use one galaxy bias parameter per redshift bin and population, with no scale dependence. Less bias parameters and other bias parameterization give similar results (see Fig. \ref{fig:fom_zmin_abs} or \citealt{2015arXiv150800035E}). We include all cross-correlations between redshift bins and different populations. We use Planck priors for all parameters except for $\gamma$ and galaxy bias.

\subsection{Figure of Merit (FoM)} \label{subsec:fom}

The Figure of Merit (FoM) for a certain parameter subspace $S$ is defined as

\begin{equation}
\text{FoM}_S \equiv \frac{1}{\sqrt{\det{\left[F^{-1}\right]_S}}},
\end{equation} 

\medskip

marginalizing over parameters not in $S$. This is a good estimator of the error for different dimensional subspaces $S$. For one parameter, then this is the inverse error (Eq. $\ref{lambdacov}$) of the parameter. For two parameters it is proportional to the inverse area included within 1-sigma error ellipse. For three parameters it is the inverse volume within 1-sigma error ellipsoid, and so on. In this paper we focus in the figures of merit defined in (\citealt{2015arXiv150203972E}):

\begin{itemize}
\item $\fomdetf$. $S=(\omega_0,\, \omega_a)$. Dark Energy Task Force (DETF) Figure of Merit (\citealt{2006astro.ph..9591A}). Inversely proportional to the error ellipse of $(\omega_0,\, \omega_a)$. The growth factor $\gamma$ is fixed.
\item $\fom$ : Equivalent to $\fomdetf$, but instead of $\gamma=0.55$ from GR, $\gamma$ is considered a free parameter and is marginalized over.
\item $\fomg$. $S= (\gamma)$. Corresponds to the inverse error of the growth parameter $\gamma$. Therefore, $\fomg=10,\, 100$ corresponds to $10\%$, $1\%$ expected error on $\gamma$. The dark energy equation of state parameters $(\omega_0, \omega_a)$ are fixed.
\item $\fomc$. $S=(\omega_0,\, \omega_a, \, \gamma)$. Combined figure of merit for $\omega_0$, $\omega_a$ and $\gamma$.
\end{itemize}

It is important to note that, when not including priors, the different FoMs scale with area $A$ in the following way

\begin{equation} \label{fomarea}
\begin{split}
&\fomdetf \propto A, \\
&\fom \propto A, \\
&\fomg \propto A^{1/2}, \\
&\fomc \propto A^{3/2}.
\end{split}
\end{equation} 

\medskip

Doubling the area would give a factor $\sim2.83$ higher $\fomc$.


\subsection{Fiducial galaxy sample} \label{subsec:fiducial}

\setlength{\parindent}{6ex}
We define two galaxy populations based on the following fiducial spectroscopic (Bright\footnote{This population definition is in correspondence with previous work such as \cite{2012MNRAS.422.2904G} and \cite{2015arXiv150203972E}}, B) population. We define a magnitude limited survey, with $i_{AB}<22.5$ as the fiducial flux limit in the i-band. The fiducial survey area is $14000$ deg$^2$. The fiducial redshift range is $0.1<z<1.25$, and the number of redshift bins is 71, with a narrow bin width of $0.01(1+z)$. Spectroscopic surveys usually have great redshift determination, so we define a Gaussian spectroscopic redshift uncertainty of $\sigma_{68}=0.001(1+z)$, much lower than the bin width.

The fiducial bias is interpolated within 4 redshift pivot points, $z=0.25, 0.43, 0.66, 1.0$, which scale with redshift in the following way,

\begin{equation}
b_B(z) = 2+2(z-0.5).
\end{equation} 

\medskip

\noindent Recall that there is one bias parameter per redshift bin and population. The fiducial redshift distribution of galaxies is characterized with the number density of objects per solid angle and redshift as

\begin{equation} \label{nofz}
\frac{dN}{d\Omega dz} = N \left(\frac{z}{z_0}\right)^\alpha \exp{\left(-\left(\frac{z}{z_0}\right)^\beta\right)},
\end{equation} 

\medskip

\noindent and is constructed by fitting a Smail type $n(z)$ (\citealt{1991ApJ...380L..47E}) to the public COSMOs photo-z sample (\citealt{2010ApJ...709..644I}). The values for $\alpha$, $\beta$ and $z_0$ in Eq. \ref{nofz} correspond exactly to the values in \citealt{2012MNRAS.422.2904G}: $z_0=0.702$, $\alpha=1.083$ and $\beta=2.628$. The normalization $N$  sets the density of galaxies per solid angle, being the fiducial density for this work $n_g=0.4$ gal/arcmin$^2$. Table \ref{paramfidsurvey} summarizes the parameters that characterize our fiducial spectroscopic survey.

\begin{table} \centering
\begin{tabular}{ l  r }
\hline
Area [deg$^2$] & 14,000 \\
Magnitude limit & $i_{AB}<22.5$ \\
Redshift range & $0.1<z<1.25$ \\
Redshift uncertainty & $0.001(1+z)$ \\
zBin width & $0.01(1+z)$ \\
Number of zbins & 71 \\  [1mm] \hdashline \\  [-2mm] 
Bias: $b(z)$ & 2+2(z-0.5) \\  [1mm]  \hdashline \\  [-2mm] 
Density [gal/arcmin$^2$] & 0.4 \\
$n(z)$ - $z_0$ & $0.702$ \\
$n(z)$ - $\alpha$ & $1.083$ \\
$n(z)$ - $\beta$ & $2.628$ \\ \hline
\end{tabular}
\caption{Parameters that describe our fiducial spectroscopic survey.}
\label{paramfidsurvey}
\end{table}

%% file: samplevar_clean.tex
\section{Sample variance cancellation} \label{sec:samplevar}

When two populations in a survey overlap in the same volume (B1xB2) one gets additional cross-correlations and covariance between them. If one is able to split one galaxy sample into two galaxy overdensities in the same area by some observable (i.e. luminosity, color), the resulting subsamples become correlated as they trace the same underlying dark matter fluctuations. As a result, using multiple tracers allow for sampling variance cancellation and can considerably improve the constraints. This multi-tracer technique was first introduced in \cite{2008.0810.0323}. 

Assume B1 and B2 are two galaxy populations, one with bias $b$ and the other with bias $\alpha b$. Their density perturbation equations  in redshift space (Eq. \ref{rsdfourierbias}) and in the linear regime are

\begin{equation} \label{b1}
\delta_{B1} (k)= (b+f\mu^2)\,\delta(k) +\epsilon_1,
\end{equation} 

\noindent
and 
\begin{equation} \label{b2}
\delta_{B2} (k)= (\alpha b+f\mu^2)\,\delta(k) + \epsilon_2 ,
\end{equation} 

\medskip

\noindent where $\mu\equiv k_\|/k$ is defined to be the cosine of the angle between the line of sight and the wavevector $\hat{k}$, and $\epsilon_i$ are stochasticity parameters that can refer to a standard shot-noise or to other random component. 

Even when having an infinite galaxy sample, there will be cosmic variance as each mode $\delta (k)$ is a random realization of a Gaussian field. However, if we have two tracers sampling the field we can average over many modes and cancel the sampling variance. To illustrate this we divide Eq.\ref{b2} over Eq.\ref{b1} (with no stochasticity) and obtain

\begin{equation} \label{cancel}
\frac{\delta_{B2}}{\delta_{B1}}= \frac{\alpha \beta^{-1}+\mu^2}{ \beta^{-1}+\mu^2},
\end{equation} 

\medskip

\noindent where $\beta\equiv f/b$, which has explicit angular dependence, but no dependence on the random field $\delta$, which allows to extract $\alpha$ and $\beta$ separately, and determine $\beta$ exactly in the absence of shot-noise. In \cite{2008.0810.0323} the authors compute an analytical example considering a pair of transverse and radial modes ($\mu=1$ and $\mu=0$), and already found that can arbitrarily improve the determination of $\beta$ with respect to the single galaxy in the limit of zero shot-noise.

Splitting one spectroscopic sample into two over the same area increases the number of observables available but also increases the number of nuisance parameters and adds shot-noise to the observables. The cosmological information coming from two overlapping tracers is correlated as well as their nuisance parameters, which manages to reduce the error on cosmological parameters (\cite{2015MNRAS.451.1553E}). Decisively, splitting optimizes the constraints by canceling the random nature in the amplitude of the modes (see Eq. \ref{cancel}).

\medskip

The way these mentioned effects propagate into the FoMs is the object of study in Section 3. In the following subsections we show the impact in our forecast of the relative bias amplitude (subsection \ref{subsec:relbias}) and the dependence on galaxy density (subsection \ref{subsec:density}) for the single and multi tracer surveys. In subsection \ref{subsec:effectsbiashalf} we show the FoMs for $\alpha=0.5$, which is the fiducial relative bias amplitude value for subsection \ref{subsec:density}, section \ref{sec:Partlyoverlapping} and Appendix \ref{appA}.

\subsection{Relative bias amplitude ($\alpha$)}  \label{subsec:relbias}  

In Fig. \ref{fig:fom_redshift_bamp_densgal_01} we show $\fomc$ ($\S$\ref{subsec:fom}) (for other FoM see Fig. \ref{fig:fom_redshift_bamp}) for the two single tracers (B1 and B2) defined in Eqs. \ref{b1} and \ref{b2}, without stochasticity, as function of the relative bias amplitude $\alpha$ (Eq. \ref{b2}). They both follow the fiducial configuration from Table \ref{paramfidsurvey} except for the $\alpha$ parameter. B2 is shown with the fiducial density and with four times less density. Furthermore, we show what happens if we merge both single tracers into one overlapping survey B1xB2, for the two density cases of B2. B1xB2 equals B1 + B2 constraints + extra correlations + extra covariance, and includes all cross redshift and cross population correlations. All lines are normalized to the B1 FoM, which does not depend on $\alpha$.

In the example considering a pair of transverse and radial modes from \cite{2008.0810.0323}, the authors find that the improvement measuring $\beta$ is proportional to

\begin{equation}
\frac{\sigma_\beta^2 (\text{1 tracer})}{\sigma_\beta^2 (\text{2 tracers})} \propto \frac{(\alpha-1)^2}{\alpha^2},
\end{equation} 

\medskip

\noindent which is minimum at $\alpha=1$. When doing the full analysis in Fig. \ref{fig:fom_redshift_bamp_densgal_01}  we take into account the whole range of $\mu$, and our results for 2 tracers (B1xB2) also show a minimum when the bias amplitudes are equal ($\alpha=1$). Note that we do not expect B1xB2 ratio to B1 to be 1 at $\alpha=1$ as B1xB2 has twice its total density and extra correlations and covariance between tracers. To remind this and avoid confusion we write the total density in parenthesis next to the population in gal/arcmin$^2$: B1xB2(0.8), B1(0.4), B2(0.4).

When increasing the bias ratio, $\alpha \neq 1$, we cancel sample variance and we quickly improve our constraints up to a factor 4 from B1(0.4) to B1xB2(0.8). If we reduce four times the density of B2(0.4) the improvement between B1(0.1) and B1xB2(0.5) is a factor $\sim 2.3$, which is lower because shot-noise is higher. For B2(0.4) the constraints are similar for $\alpha < 1$ (lower bias amplitude), and get worse with $\alpha > 1$ (more bias). Here two effects overlap: RSD effect becomes more important with lower bias which has a great impact in $\gamma$ constraints, whereas a higher bias increases the amplitude of the correlations, which weakens the impact of shot noise, and in particular improves the $\omega$ constraints. For this reason, reducing B2 density has a larger impact on lower bias both in B2 and B1xB2 as it mitigates the benefits from RSD, and also as the signal from correlations is lower then shot noise is more predominant.

For a detailed study of the impact of RSD and BAO with bias and $\alpha$ in all FoMs see $\S$\ref{subsubsec:reds} to \ref{subsubsec:rsdbao}.

\xfigure{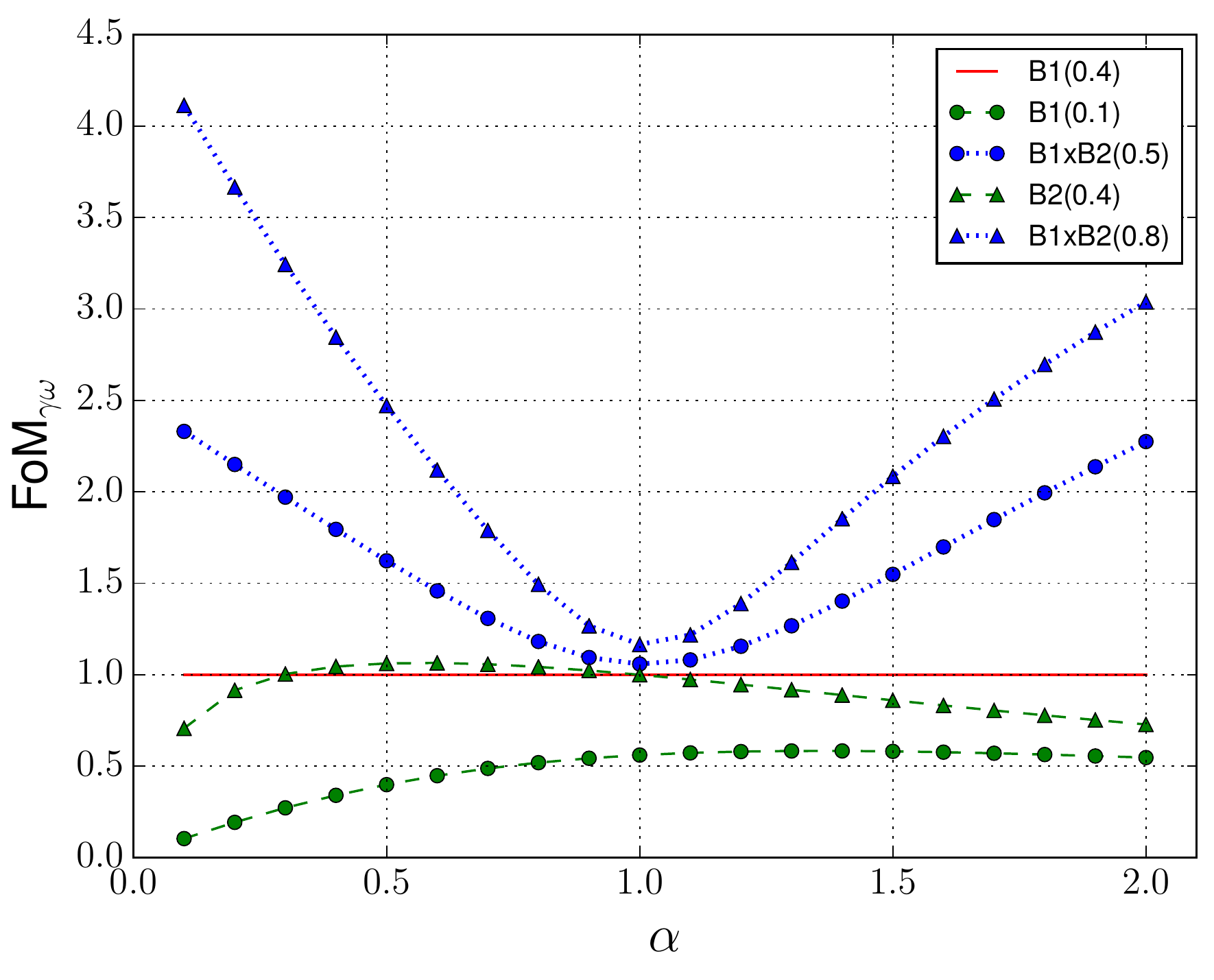}
{$\fomc$ dependence on the relative bias amplitude $\alpha$ (Eq.\ref{b2}) for the fiducial density (circles) compared to when B2 density is four times lower, 0.1 gal/arcmin$^2$, (triangles). The blue dotted lines correspond to B1xB2(0.8 or 0.5), the green dashed to B2(0.4 or 0.1) and the red solid to B1(0.4), where the values inside parenthesis indicate the total density of each population in gal/arcmin$^2$. All lines are normalized to the B1(0.4) forecast.}
{fig:fom_redshift_bamp_densgal_01}

\subsection{Fiducial model ($\alpha=0.5$)}  \label{subsec:effectsbiashalf}  

In this subsection we study several effects fixing $\alpha=0.5$, which will be the fiducial value for the relative bias amplitude in the following subsections, except for section \ref{sec:clfhm}. Table \ref{b1b2allb2amp05} presents four tabulars, one for each FoM, with the two single population cases (B1(0.4), B2(0.4)) and the multitracer case (B1xB2(0.8)) for the rows. In the columns we present the fiducial case (labeled `Fiducial') and the impact of some physical effects, like fixing bias (`xBias'), computing correlations in real space (`No RSD'), not including BAO wiggles (`No BAO') and combinations of these.

\begin{table*} \centering
  \begin{tabular}{ r  l  r  r  r  r  r  r  }
    \cmidrule{2-8}
    &&  \hspace{0.2cm}Fiducial & xBias  & No RSD & No BAO & No RSD-xBias & No BAO-xBias\\ \cmidrule{2-8} 
    \multirow{3}{*}{$10^{-3}$ FoM $\gamma\omega$:}
    &B1xB2 & 13.7 & 117 & 1.61 & 9.25 & 41.4 & 64.2 \\
    &B2 & 5.88 & 36.1  & 0.62 & 4.37 & 16.7 & 24.7 \\
    &B1 & 5.53 & 45.4  & 1.45 & 3.78 & 37.9 & 31.6 \\ \cmidrule{2-8}
    \multirow{3}{*}{FoM $\gamma$:}
    &B1xB2 & 62 & 190 & 9.9 & 58  & 105 & 143 \\
    &B2 & 51 & 152 & 7.6 & 49  & 78 & 121 \\
    &B1 & 38 & 147 & 9.6 & 38  & 102 & 133 \\ \cmidrule{2-8}
    \multirow{3}{*}{FoM $\omega$:}
    &B1xB2 & 221 & 615 & 163 & 160  & 395 & 450 \\
    &B2 & 116 & 238 & 82 & 90  & 212 & 204 \\
    &B1 & 147 & 310 & 152 & 100 & 373 & 238 \\ \cmidrule{2-8}
    \multirow{3}{*}{FoM DETF:}
    &B1xB2 & 237 & 875 & 209 & 171  & 841 & 787 \\
    &B2 & 129 & 513 & 106 & 104 & 479 & 422 \\
    &B1 & 180 & 801 & 196 & 137 & 797 & 696 \\ \cmidrule{2-8}
 
   \end{tabular}
\caption{Sample variance cancellation for multitracing B1xB2(0.8) of two spectroscopic populations, B1(0.4) and B2(0.4), where the values inside parenthesis indicate the total density of each population in gal/arcmin$^2$. The relative bias amplitude between both populations is set to $\alpha =0.5$. Each column show the impact of removing different effects, while rows show the single and overlapping population cases.}
\label{b1b2allb2amp05}
\end{table*}

\xxfigure{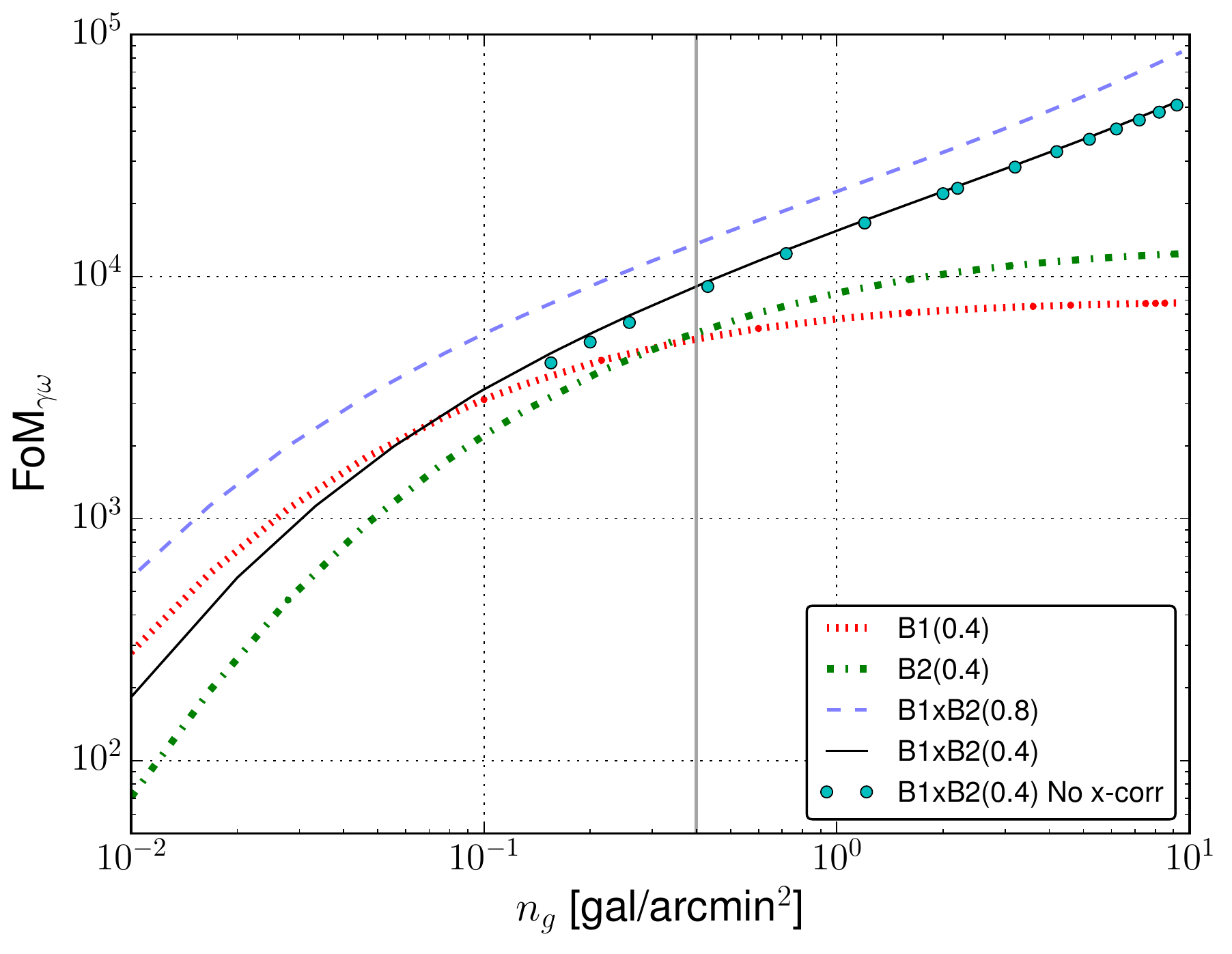}{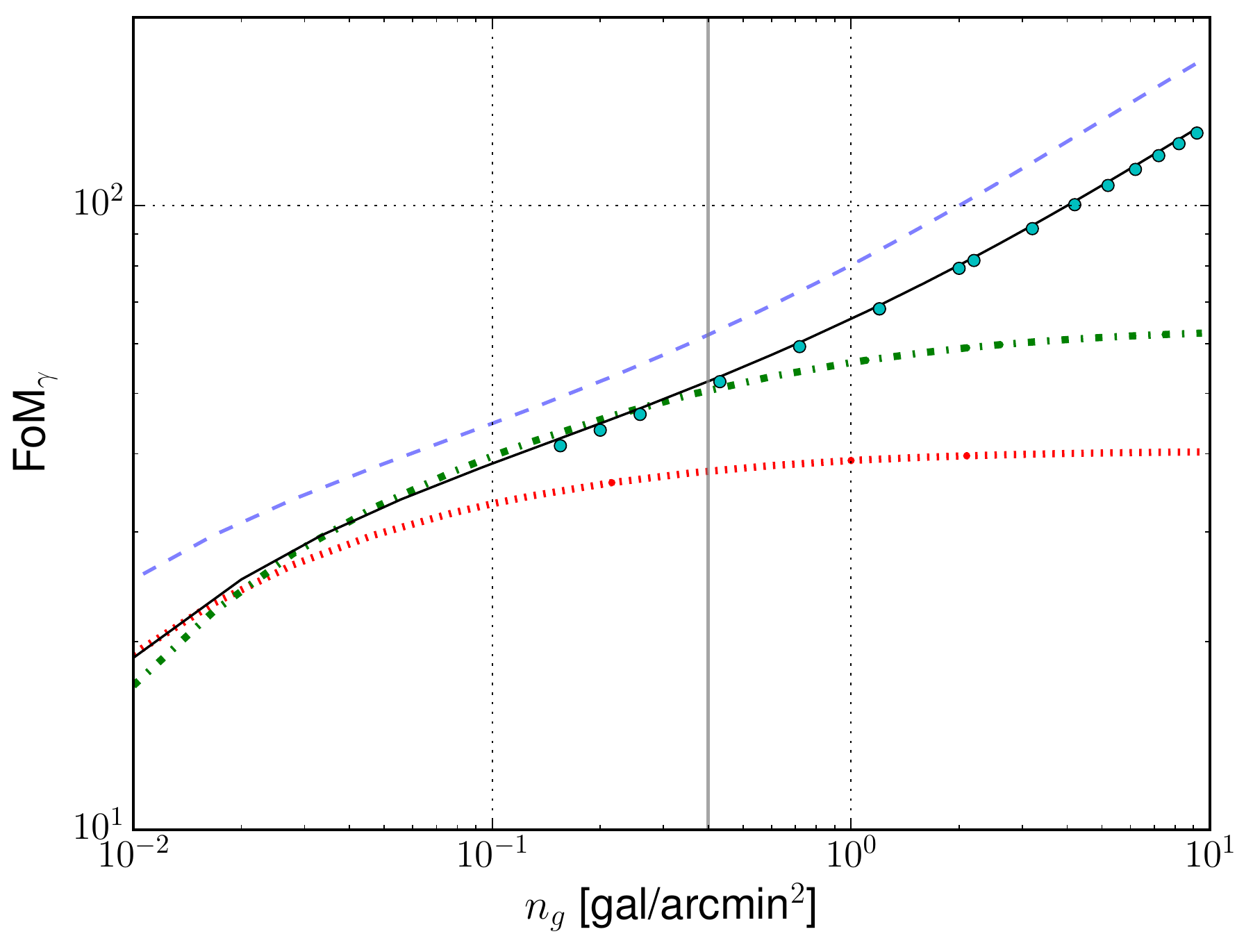}
{Impact of spectroscopic galaxy density on the constraints. The relative bias amplitude is fixed at $\alpha=0.5$. The red dotted B1(0.4) and green dot-dashed B2(0.4) lines correspond to the single tracers. B1xB2(0.8) (blue dashed line) is the overlapping survey of merging B1(0.4) and B2(0.4), and thus has double the density of each alone. B1xB2(0.4) (black solid) has the same total density as B1(0.4) or B2(0.4) and is identical to B1xB2(0.8) except its tracers have halve the density. The cyan dots show B1xB2(0.4) without cross correlations between B1(0.2) and B2(0.2), which is equivalent to adding the auto correlations from each B1(0.2) and B2(0.2) population alone plus the covariance between them. The vertical line shows the fiducial density, $0.4$ gal/arcmin$^2$.}
{fig:foms_densityX}

Looking first at the `Fiducial' column, one sees how the multitracer case has better constraints than the single tracer cases, for all FoMs, due to sample variance cancellations. Comparing to the best single tracer, there is a 133\% improvement for $\fomc$, 23\% for $\fomg$, 50\% for $\fom$ and 32\% for $\fomdetf$. 

Galaxy bias can be fixed from lensing surveys and its cross-correlations with galaxy clustering (see \citealt{2011MNRAS.416.3009B},  \citealt{2012MNRAS.422.1045C}, \citealt{2012MNRAS.422.2904G},  \citealt{2015arXiv150203972E}). Fixing bias greatly improves the constraints as it breaks strong degeneracies, but the gains from sample variance cancellations are still present, which shows that they are not caused by measuring bias. RSD allow to measure galaxy bias and the growth separately, but not the random nature of the fluctuations, so fixing bias will not break the degeneracy with the rms amplitude of fluctuations, but multiple tracers will. When removing redshift space distortions (`No RSD'), sample variance cancellations are no longer possible, and the gain for B1xB2(0.8)/B1(0.4) is much lower. Also, without RSD, our ability to measure $\gamma$ drops, which translates in a much lower FoMs. Not including BAO measurements reduce the FoMs, affecting more the $\omega$ constraints while having little impact on $\fomg$ (see $\S$\ref{appA} for a discussion of the impact of RSD and BAO). We have also checked the effect of weak lensing magnification using the magnification slopes given in \cite{2015arXiv150203972E}. We find that they contribute less than 0.5\%.

\subsection{Galaxy density}  \label{subsec:density}  

The auto-correlations for a redshift bin include a shot-noise term (see $\S$\ref{subsec:corr}) due to the discrete nature of the observable (galaxy counts), which depend on the galaxy density. Previously in the introduction of section $\S$\ref{sec:samplevar} we have discussed that multiple tracers in redshift space can cancel sampling variance, and then our ability to improve our constraints is only limited to the signal-to-noise of the tracers (except when including bias stochasticity). Therefore, if there is no bias stochasticity, by increasing the survey density we can improve our cosmology constraints as much as we want. However, surveys usually have a fixed exposure time, so increasing survey density requires going deeper (longer exposures), which results in a smaller survey area. In this subsection we do not study this trade off between galaxy density and area, but increase galaxy densities for a fixed area. Moreover, spectroscopic surveys are characterized by having very good redshift determination since it has spectra where one can locate the emission lines, but it requires to take longer exposure times which results into lower densities. 

Fig. \ref{fig:foms_densityX} shows how $\fomc$ and $\fomg$ depend on galaxy density. B1(0.4) and B2(0.4) correspond to the single tracer surveys. The blue line B1xB2(0.8) is a multiple tracer survey which merges the single tracer surveys B1(0.4) and B2(0.4) over the same area, as the multitracer surveys studied in Fig. \ref{fig:fom_redshift_bamp_densgal_01} and Table \ref{b1b2allb2amp05}. Therefore, it has double of the density of one single tracer alone, and the x-axis refers to the density of one of the subsamples of the survey. On the other hand, the black line B1xB2(0.4) studies the constraints when splitting one single tracer like B1(0.4) into two, keeping the total density, and thus the density of each subsample is reduced by half. Therefore, when comparing to the single tracers the black line addresses the gains from covariance and cross-correlations but adding shot-noise in the subsamples, and the blue line adds the gains from extra density from merging B1(0.4) with B2(0.4). Note that in a real survey we are interested in the gains coming from splitting into two subsamples like B1xB2(0.4), while B1xB2(0.8) shows the combined constraints the way B1(0.4) and B2(0.4) are fiducially defined.

As expected, the single tracer results flatten out at the high density limit and saturate. For multiple tracers there is sample variance cancellation and the constraints improve beyond the single tracer noiseless limit. At lower densities we observe that B1xB2(0.4) and B1(0.4) lines cross. When shot-noise is already high we do not expect further splitting to improve the constraints. Moreover, B1xB2(0.8) shows better constraints than B1xB2(0.4) as the higher density reduces shot-noise. For the single tracers, in $\fomg$ the constraints are similar for B1(0.4) and B2(0.4) at low densities, while we observe a clear difference (due to bias) between them on the noiseless limit. Lower bias populations (B2) get better constraints because in $\fomg$ RSD is vital for breaking degeneracies and is enhanced with a lower bias amplitude (see $\S$\ref{subsubsec:reds} for details), but when shot noise becomes dominant then this effect disappears. On the other hand, in $\fomc$ we observe that B1(0.4) and B2(0.4) lines cross, as for the $\omega$ constraints a lower bias gives lower constraints in general, but this effect is more noticeable with high noise, as in that case having more signal is more relevant, while in the noiseless limit the constraints are similar for different bias (see, for example, right panel in Fig. \ref{fig:fom_zrange_biaslope}). 

The cyan dots correspond to removing cross correlations between B1(0.2) and B2(0.2) in  B1xB2(0.4). This is equivalent to adding the correlations (transverse and radial within redshift bins), of each population B1(0.4) and B2(0.4) (B1+B2) plus the same sky covariance. It shows the relative importance of covariance between the tracers and the additional cross correlations in the gains that we are observing. We find that there is only a tiny contribution from cross-correlations ($<2\%$ at high density), which shows that the multiple tracer improvement comes mainly from sample variance cancellations.

From now on we only show the multitracer B1xB2 with the same total density as the single tracers.

\xfigure{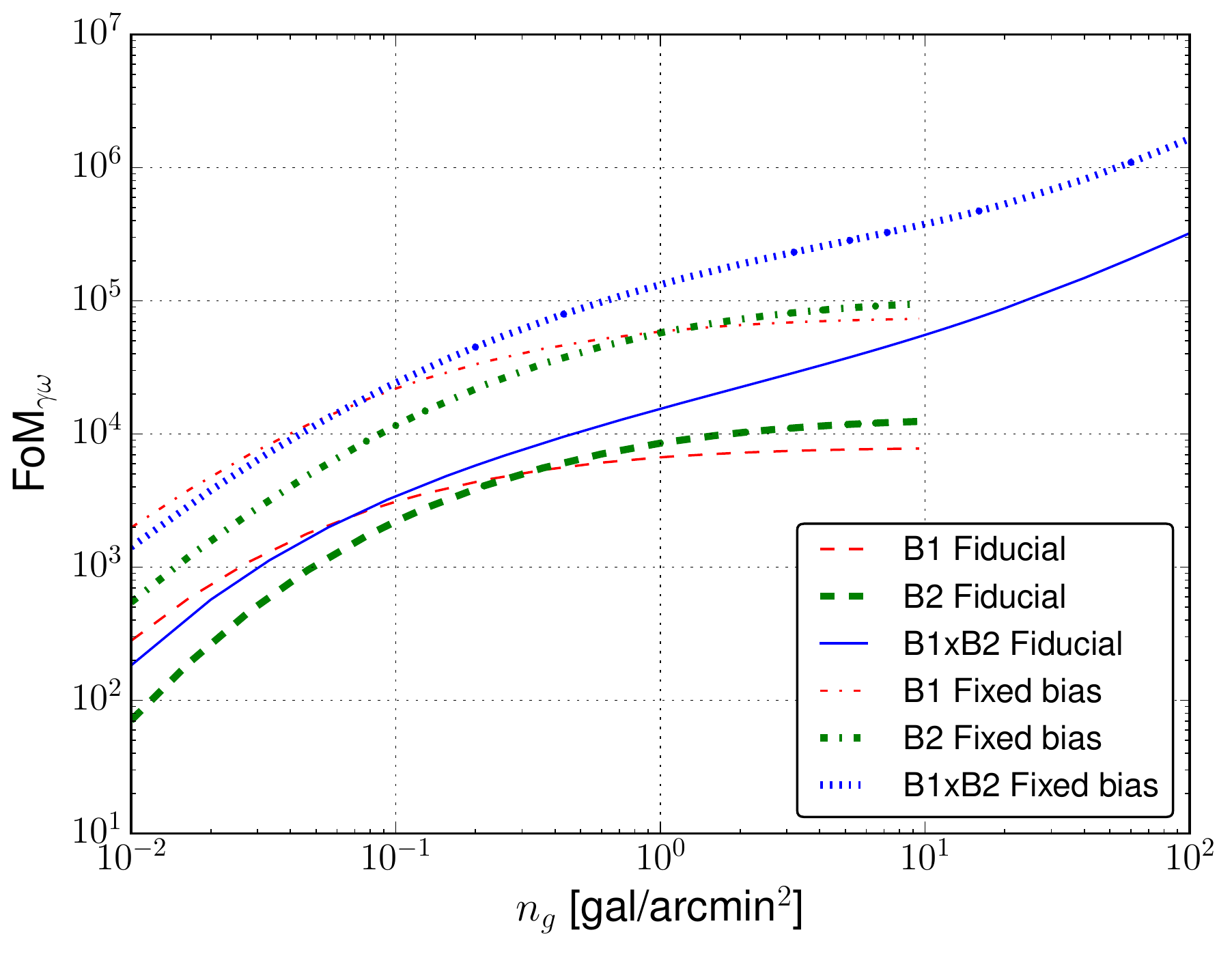}
{Shows the constraints of $\fomc$ for the free and fixed bias case, for the B1(0.4), B2(0.4) and B1xB2(0.4) surveys. Free bias means marginalizing over the bias parameters, and corresponds to the fiducial forecast (Labeled as 'Fiducial').
}
{fig:foms_xbias}

\subsubsection{Fixing bias} \label{subsubsec:xbias} 

Fig. \ref{fig:foms_xbias} shows the constraints from B1xB2(0.4), B1(0.4) and B2(0.4) when fixing bias compared to the free bias case (free bias means marginalizing over the bias parameters). When we fix bias we break strong degeneracies and the constraints improve by an order of magnitude. We find that not all improvement comes from measuring bias, as we find similar relative gains with the fixed and free bias cases. The fact that for B1xB2(0.4) the free and fixed bias lines approach comes from the $\gamma$ constraints, while for the $\omega$ constraints the difference is rather flat (not shown). Note that some extreme density values are shown, which are included to study the potential gains from a theoretical point of view.

%% file: hod_clean.tex
\section{Relation between bias and density}  \label{sec:clfhm} 

In the last subsections we have studied the constraints dependence on galaxy density and relative bias amplitude when splitting one spectroscopic population into two. When splitting by luminosity or absolute magnitude, brighter galaxies tend to live in more massive haloes, which tend to be more biased and less abundant. Therefore, there is a relation between galaxy density and bias. In the other sections we fix $\alpha=0.5$ and ignore this relation to understand different physical effects from a theoretical point. To account for this effect, in this subsection we model galaxy bias using a conditional luminosity function (CLF) fitted to SDSS data from \cite{2013MNRAS.430..767C} combined with a halo model (HM). The CLF determines how galaxies with a given luminosity populate dark matter haloes of different mass, $\Phi(L|M)$, while the HM set the abundance of dark matter haloes of a certain mass, $n(M,z)$. Using this modeling we define a magnitude limited survey $18 < r_{AB} < 23$ and we are able to determine the abundance of galaxies and galaxy bias as a function of redshift, halo mass or galaxy luminosity. To define the apparent limited survey we only consider luminosities in redshift such that $r_{AB}(L,z) \in [18,23]$, since \cite{2013MNRAS.430..767C} fit the HOD model using the SDSS r-band data.

\subsection{Conditional Luminosity Function}

The conditional luminosity function from \cite{2013MNRAS.430..767C} has two separate descriptions for the central and satellite galaxies:

\begin{equation} \label{eq:clf1}
\begin{split}
\Phi(L|M) &= \Phi_c(L|M) + \Phi_s(L|M) ,\\
\Phi_c(L|M) \,dL &= \frac{\log{e}}{\sqrt{2\pi} \sigma_c} \exp\left[{-\frac{(\log{L}-\log{L_c})^2}{2\sigma_c^2}}\right]\frac{dL}{L},\\
\Phi_s(L|M) \, dL &= \phi_s^{*}\left(\frac{L}{L_s^{*}}\right)^{\alpha_s+1} \exp\left[-\left(\frac{L}{L^{*}_s}\right)^2\right] \frac{dL}{L},
\end{split}
\end{equation}

\noindent
where $\log$ is the 10-based logarithm and  $L_c$, $\sigma_c$, $\phi_s^{*}$, $\alpha_s$ and $L_s^{*}$ are all function of halo mass $M$,
\begin{equation} \label{eq:clf2}
\begin{split}
L_c(M) &= L_0\frac{(M/M_1)^{\gamma_1}}{[1+(M/M_1)]^{\gamma_1-\gamma_2}}, \\
L_s^{*}(M) &= 0.562\, L_c(M),\\
\alpha_s(M)&= \alpha_s, \\
\log{[\phi_s^{*}(M)]} &= b_0 + b_1(\log{M_{12}})+ b_2(\log{M_{12}})^2.
\end{split}
\end{equation}

\medskip

\noindent For the total set of CLF parameters we use the median of the marginalized posterior distribution given in \cite{2013MNRAS.430..767C} for their fiducial model.

\subsection{Halo Model}

The comoving number density of haloes per unit halo mass can be well described (\citealt{1974ApJ...187..425P}, \citealt{1999MNRAS.308..119S}) by
\begin{equation} \label{eq:massf}
\frac{dn_h}{dM} = f(\sigma) \frac{\rho_m}{M^2} \frac{d\ln{\sigma^{-1}}}{d\ln{M}},
\end{equation}

\medskip

\noindent where $\rho_m$ is the mean density of the universe and $\sigma^2(M,z)$ the density variance smoothed in a top hat sphere at some radius $R(M)=(3M/4\pi \rho_m)^{1/3}$, 

\begin{equation}
\sigma^2(M,z) = \frac{D^2(z)}{2\pi^2} \int dk \, k^2\, P(k) \left|W(kR)\right|^2 ,
\end{equation}

\medskip

\noindent where $W(x)=3j_1(x)/x$. For the differential mass function $f(\sigma, z)$ we use the fit to the MICE simulation from \cite{2010MNRAS.403.1353C},

\begin{equation} \label{eq:fmice}
f(\sigma, z) = A(z) \left[ \sigma^{-a(z)}+b(z)\right] \exp{\left[-\frac{c(z)}{\sigma^2}\right]}
\end{equation}

\medskip

\noindent with $A(z) = 0.58(1+z)^{-0.13}$, $a(z) = 1.37(1+z)^{-0.15}$, $b(z) =0.3(1+z)^{-0.084}$, $c(z) = 1.036(1+z)^{-0.024}$. We define the halo mass function in arcmin$^2$ units as

\begin{equation}
n_h(M,z) \equiv \frac{dN_h/ dM}{d\Omega \,dz} = \left(\frac{\pi}{10800}\right)^2 \frac{c\,\chi^2(z)}{H(z)}  \frac{dn_h}{dM}(z).
\end{equation}

\medskip

\noindent To model halo bias function we use the fitting function from \cite{2010ApJ...724..878T},

\begin{equation} \label{eq:biasfun}
b_h(M, z) = 1 - \frac{A(z)\, \nu^{a(z)}}{\nu^{a(z)} +\delta_c^{a(z)} } + B(z)\nu^{b(z)} + C(z) \nu^{c(z)}
\end{equation}

\medskip

\noindent where $\nu\equiv \delta_c/\sigma(M,z)$,  $\delta_c\simeq1.686$ is the linear density collapse, and where we use the parameter values from Table 2 with $\Delta=200$ from the same paper (see also \citealt{2015MNRAS.450.1674H} for other values).
\subsection{Splitting methods}

\noindent Once the halo mass function and the halo bias function are specified we can determine the galaxy number density and galaxy bias for an apparent limited survey. The average number of galaxies of a given halo mass with $L_1 < L < L_2$ is

\begin{equation} \label{phimz}
\Phi(M,z) = \int^{L_2(z)}_{L_1(z)} \Phi(L|M)\, dL,
\end{equation}

\noindent and the number density of galaxies per unit redshift is
\begin{equation} \label{ngz}
\bar{n}(z)= \int_{M_{min}}^{M_{max}} \Phi(M,z)\, n_h(M,z)\, dM,
\end{equation}

\noindent while the corresponding mean galaxy bias is
\begin{equation} \label{hodbias}
\bar{b}(z) = \int dM \,b_h(M, z)  \, \Phi(M,z)\, n_h(M,z)/\bar{n}_g(z).
\end{equation}

\xxfigure{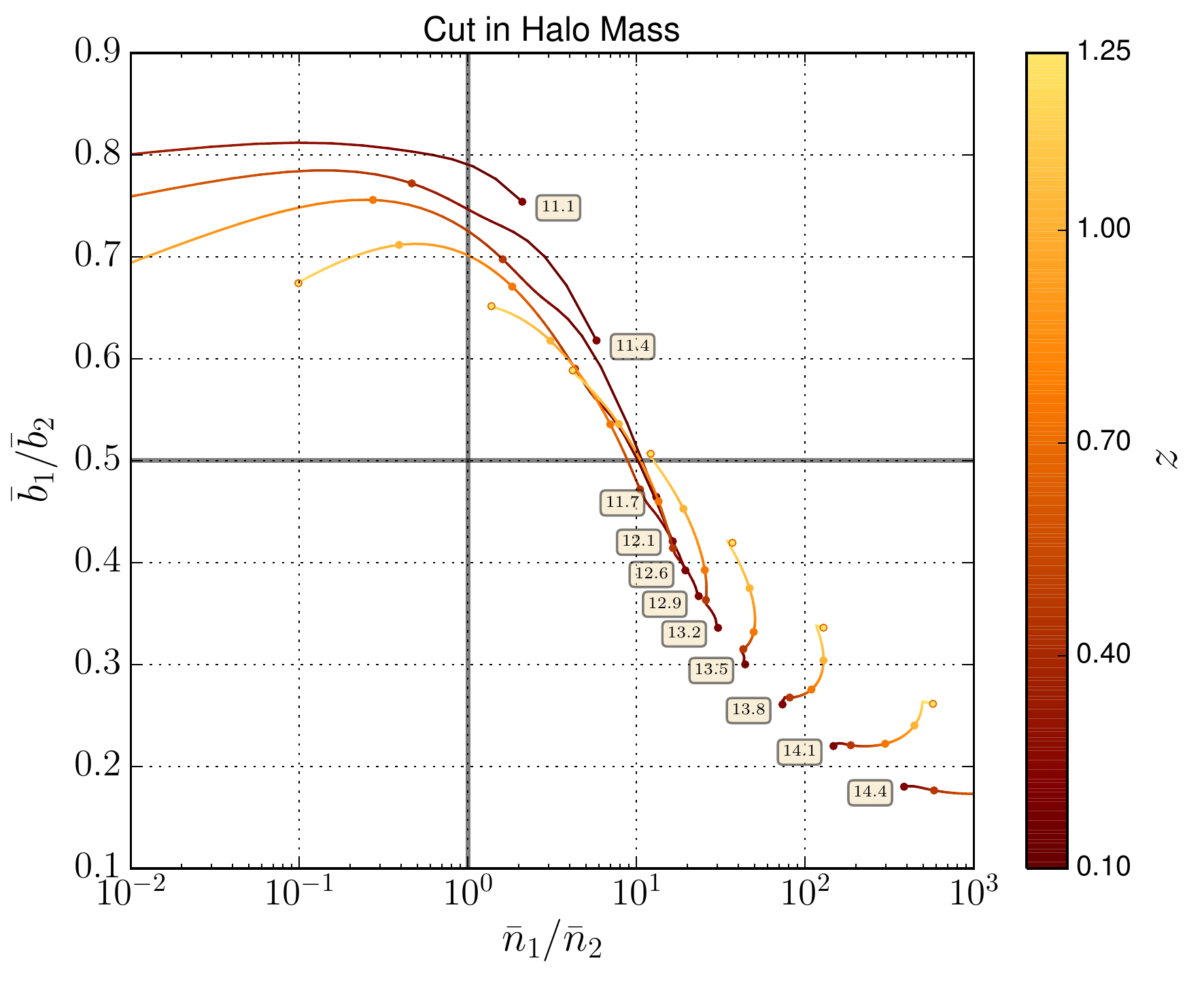}{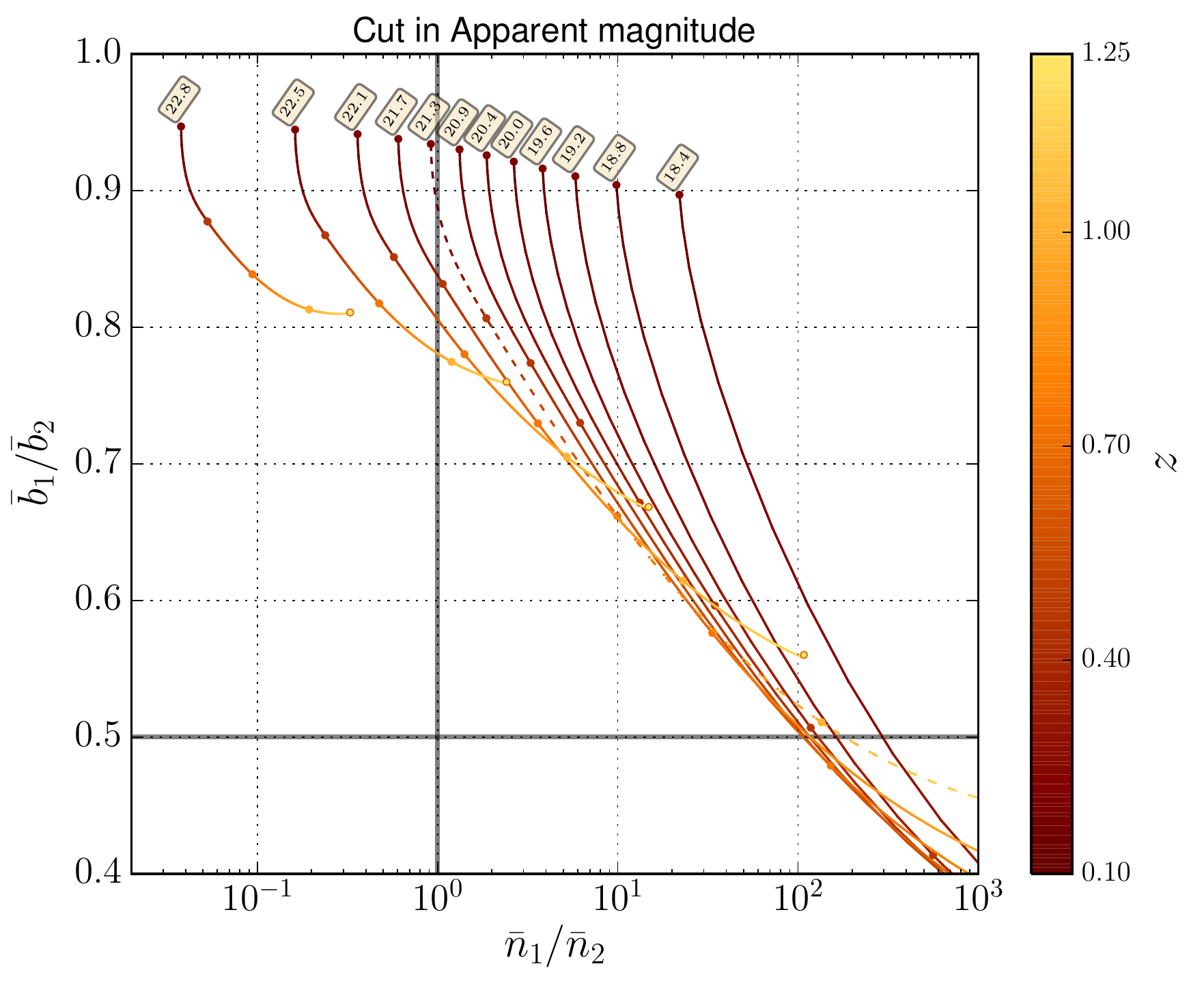}
{Density versus bias ratios between the two subsamples. The top panel shows the split with halo mass, $M_{cut}$, while the bottom panel shows a split with r-band magnitude, $r_{cut}$. Each line corresponds to a given $M_{cut}$/$r_{cut}$, which value is indicated in a box next to the start of each. The colorbar shows the redshift evolution for each line. There are 5 dots in each line indicating the position of $z=0.1,\,0.4,\,0.7,\,1.0,\,1.25$. The dashed line shows the case where $\fomc$ is maximum (see the details in the text). }
{clf_debug}

\xfigurefull{}{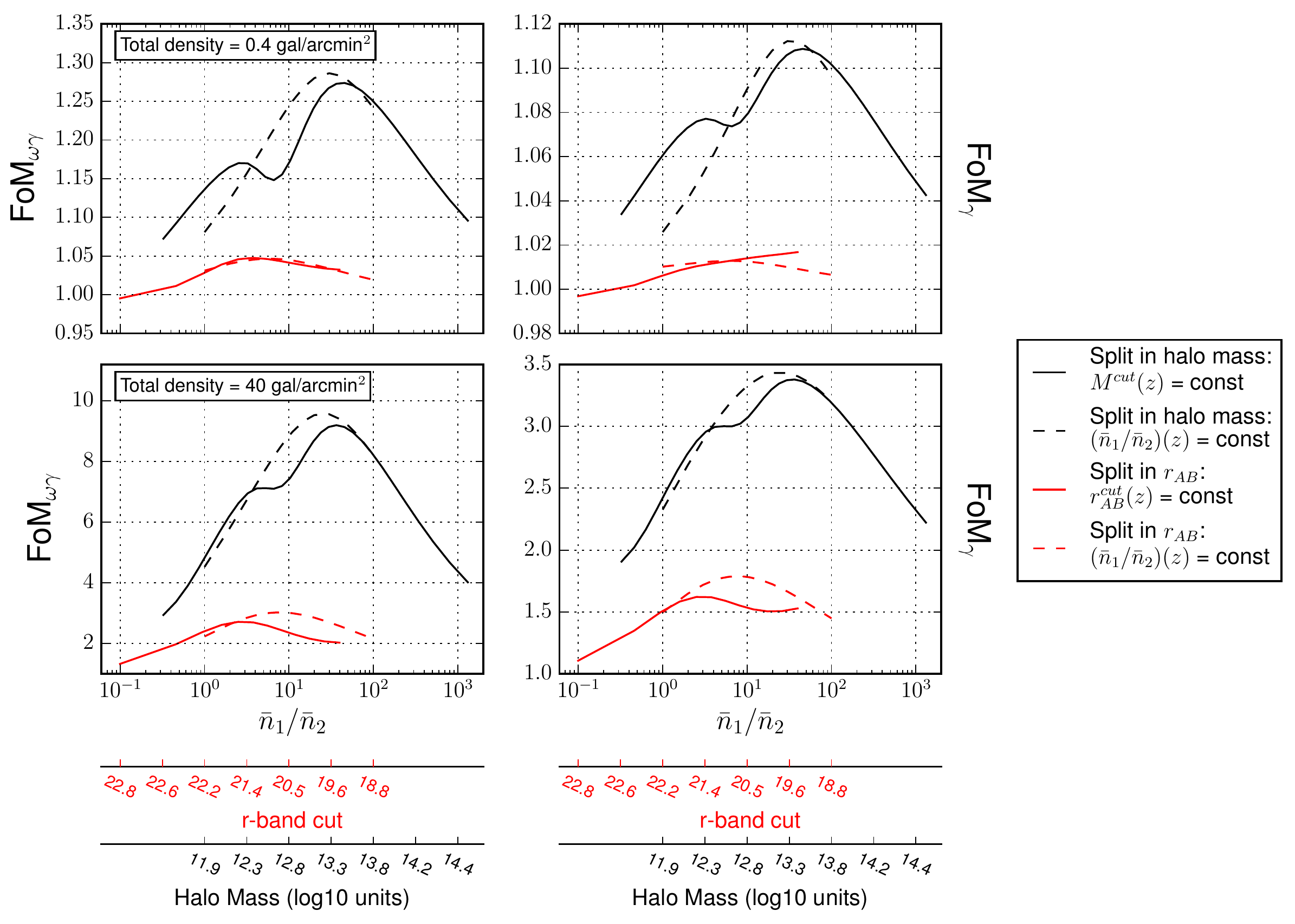}
{$\fomc$ (left panels) and $\fomc$ (right panels) when using a CLF $\Phi(L|M)$ and HM models to build an r-band  limited magnitude survey, $r_{AB}=[18,23]$. Two splitting methods are shown, splitting in halo mass, $M_{cut}$, and splitting in r-band, $r_{cut}$. Two cases are studied for each method, splitting with constant $M_{cut}$/$r_{cut}$ in redshift and splitting with constant density ratio in redshift, $\bar{n}_{1}(z)/\bar{n}_{2}(z) \propto const$. The top panels have a total density of $0.4$ gal/arcmin$^2$, while the bottom panels have $40$ gal/arcmin$^2$. The x-axis shows the density ratio between the two subsamples, and the two twin axis show the correspondence of this ratio to a given constant $M_{cut}$ and $r_{cut}$ in redshift. All lines are normalized to the FoM when not splitting the galaxy sample.}
{fomall_clf}

\noindent Here we define $L_1(z)$ and $L_2(z)$ such that $r_{AB} (L_2(z)) = 18$ and $r_{AB} (L_1(z)) = 23$. We integrate between $M_{min} = 10$ and $M_{max} = 15$ in $\log{ [M/M_\odot h^{-1}]}$ units and consider $\Phi(L|M) =0$ outside of this boundaries. To split the survey into two subsamples we consider two methods:
\begin{itemize}
\item Splitting by halo mass: split the spectroscopic sample introducing a $M_{cut}$ in Eqs. \ref{ngz}-\ref{hodbias}  which defines two
populations, B1 with $M_{min} < M < M_{cut}$ and B2 with $M_{cut} < M < M_{max}$.
\item Splitting by apparent magnitude: split the spectroscopic sample introducing an $L_{cut}(z)$ in Eq.\ref{phimz} which defines two populations, B1 with $L_{1} < L < L_{cut}$ and B2 with $L_{cut} < L < L_{2}$. Notice that $r_{AB}(L_{cut}(z)) = r_{cut}$.
\end{itemize}

\medskip

\noindent Within this two methods we consider two cases, one in which the cutting variable ($M_{cut}$ and $r_{cut}$) is the same for all redshifts. The another case fix the density ratio (i.e. $\bar{n}_{1}(z)/\bar{n}_{2}(z) = const.$) as a function of redshift by fitting the $M_{cut}(z)$ and $r_{cut}(z)$ which produces the corresponding density ratio. This results in a total of four different forecasts. Notice that fixing the density ratio cutting in apparent magnitude $r_{cut}(z)$ or absolute magnitude (luminosity) $L_{cut}(z)$ is the same.

Fig. \ref{fomall_clf} shows the four cases that have just been described for $\fomc$ in the left panels and $\fomg$ in the right panels. Two density cases are studied, $0.4$ gal/arcmin$^2$ (top panels) and $40$ gal/arcmin$^2$ (bottom panels). The x-axis shows the density ratio between the two subsamples for each case, while the two twin axis show the correspondence of this density ratio to the cutting variable (halo mass and apparent r-band magnitude) for the two cases in which the cutting variable is constant in redshift. All lines have been normalized to the FoM when not splitting the galaxy sample.

Fig. \ref{fomall_clf} shows that a split of galaxies using the halo mass gives a better improvement in the constraints than splitting with apparent magnitude. Splitting with halo mass improves up to a factor 1.27 in $\fomc$ with low density (top left panel) while splitting with an r-band cut gives a factor 1.05. The peaks are found at halo mass $M_{cut}\simeq13.5$ ($\log{ [M/M_\odot h^{-1}]}$) and $r_{cut}\simeq21.3$. Forcing the density ratio between the subsamples to be the same in redshift (labelled as cut with constant density in Fig. \ref{fomall_clf}) slightly improves the constraints to a factor 1.29 for a cut in halo mass and leaves it near the same for an r-band split.  When using a denser population (bottom left) the improvement raises to a factor 9.2 in $\fomc$ for a halo mass split and a factor 2.7 for r-band split. When fixing the density ratio the factors are 9.6 and 3.0, respectively. The maximum gains are obtained for $\bar{n}_{1}/\bar{n}_{2} \sim 7$ when cutting in r-band and $\bar{n}_{1}/\bar{n}_{2} \sim 30$ when cutting in mass. In practice, one does not need to know the mass or the r-band, but only to have an observational proxy that allows to rank the galaxies to allow the sample split (e.g., richness in the case of halo mass). For $\fomg$ and low density (top right) the factors are 1.11 and 1.02 for $M_{cut}\simeq 13.5$ and $r_{cut}\simeq19.4$, although for the r-band cut the maximum would be found at brighter cuts which were numerically unstable. For a denser survey (bottom right), when fixing the density ratio, the constraints improve up to a factor 3.43 for halo mass and 1.79 for r-band.

When splitting a population into two subsamples one want to maximize the bias difference in redshift between them while keeping their densities as similar as possible in order to maximize the FoM. To do so, we would like to have a quantity that increase monotonically with bias with small scatter. Halo mass is such a quantity and so it maximizes the FoM. Splitting in apparent magnitude gives a distribution in halo mass, $\Phi(L|M)$, reducing the bias difference. 

Fig. \ref{clf_debug} shows the density-bias ratio evolution in redshift for different cut values when cutting in halo mass (top panel) and r-band magnitude (bottom panel). The dots in the figure show the position of the 5 ticks from the colorbar ($z=0.1,\,0.4,\,0.7,\,1.0,\,1.25$). For a halo mass cut the bias difference is low when splitting at low halo masses as bias evolves linearly in that regime and the abundance of galaxies overweights that region in front of the high biased one. Cutting at higher masses results into an increasingly greater bias difference, but also makes a more uneven density split. The maximum in $\fomc$ is found at $M_{cut}\simeq13.5$, which has a similar density ratio in redshift $\bar{n}_{1}/\bar{n}_{2}= 40\sim50$ and a bias difference of $\alpha=\bar{b}_{1}/\bar{b}_{2}= 0.3\sim0.4$. 

When splitting with apparent magnitude (Fig. \ref{clf_debug}, lower panel) the density ratios quickly span over large ranges in redshift when the bias difference increases, which limits the amount of improvement. For most magnitude cuts an important part of the distribution is very unevenly splitted, which increase the shot-noise. Furthermore, at a density ratio of $40\sim50$, (i.e. the peak with a halo mass cut in Fig.\ref{fomall_clf}), there is no magnitude cut at any redshift which produces an $\alpha\lesssim0.55$, which is a factor $1.4\sim1.8$ less bias difference than in the halo mass situation. In the high density case we are not shot-noise dominated and thus the improvement goes from a marginal 5\% to a 3 times better $\fomc$.

In addition, Fig. \ref{fomall_clf} shows a relative minimum at $M_{cut}\simeq12.6$ and a relative maximum at $M_{cut}\simeq12.1$  for the halo mass cuts at lower density cases, in both $\fomc$ and $\fomg$. Fig. \ref{clf_debug}
shows that although $M_{cut}\simeq12.6$ has a $10\%\sim15\%$ greater bias difference depending on redshift it has a more uneven density split. A cut in $M_{cut}\simeq12.1$ gives a density ratio in redshift which extends over $\bar{n}_{1}/\bar{n}_{2}\sim [0.1, 16]$, with some cuts in redshift being close/equal to a density ratio of unity, which maximally reduces shot noise,  whereas a cut in $M_{cut}\simeq12.6$ results in $\bar{n}_{1}/\bar{n}_{2}\sim [1.3, 20]$. The increment in bias difference does not compensate the induced shot noise. With higher density (Fig. \ref{fomall_clf} lower panels) shot-noise has a lower impact and the relative minimum disappears resulting in a flattened region instead.

Moreover, we have split in absolute magnitude (not shown) by fixing the luminosity cut $L_{cut}$ as a function of redshift. The FoM were worse than with an apparent magnitude cut, and in most cases worse than not splitting the sample at all. Having a magnitude limited survey gives an incompletness of luminosity in redshift, meaning that several redshift ranges have very few galaxies or no galaxies at all, which introduces large amounts of shot-noise.

%% file: partoverlap_clean.tex
\setlength{\parindent}{6ex}
\section{Partly overlapping redshift bins} \label{sec:Partlyoverlapping} 

 In Fig. \ref{fig:fom_zmin_abs} we show the effect of having partly overlapping redshift bins between two spectroscopic populations (B1xB2) by shifting the beginning of the redshift range $z_{min}$ of one of the populations (B1) while keeping the other fixed. This shifts all the B1 redshift bins with respect to the B2 ones and determines the amount of overlap between them. The total density of both the multitracer and single tracers is equal to 0.4 gal/arcmin$^2$.  In Fig. \ref{fig:fom_zmin_abs}, the panels on the left show the FoM normalized to the fully overlapping bins value (i.e. normalized to the B1 $z_{min}=0.1$ or $0\, \Delta z$ shift value of the FoMs) for $\fomc$, $\fomg$ and $\fom$, while the panels on the right show the absolute values. The fiducial forecast line (red solid) shows oscillations that are minimum at the edges of the fiducial binning (marked by vertical grey lines on the plots) and are maximum when the redshift bins half overlap with each other (when B1 bins start in the middle of a B2 bin and viceversa). In the fiducial forecast we parametrize bias with one parameter per redshift bin and tracer. The black dashed lines show an alternative bias parameterization which parametrize the bias with four redshift pivot points $z_i\in[0.25, \,0.43, \,0.66,\, 1.0]$ and linearly interpolate between them. We find similar constraints from both bias parameterizations and this shows that the gain does not artificially come from the choice of bias parameterization.
 
When bins half overlap with each other (when B1 bins start in the middle of a B2 bin and viceversa) the gain is maximum, a factor 1.33 for $\fomc$, 1.06 for $\fomg$, 1.26 for $\fom$ and 1.33 for $\fomdetf$ (not shown). Having partially overlapping bins induces an effective thinner binning that allows to probe smaller scales which improve constraints. Most of this improvement comes from the cross correlations between both populations, as the smaller scales information comes mostly from cross-correlating with the shifted bins. When removing them (red solid to pink dash-dash-dot line) the gain factors at the peaks reduce to 1.07 for $\fomc$, 1.00 for $\fomg$, 1.07 for $\fom$ and 1.12 for $\fomdetf$, and for $\fomg$ (center left panel) shifting bins even leads to worse constraints. When B1 $z_{min}$ starts exactly at the second bin of B2 the constraints drop as all bins perfectly overlap again, but with the forecast having one less bin the FoMs are slightly lower compared to the fiducial forecast. The effect of removing the first bins does not reduce much the FoMs as the first bins are often removed from cutting in $k$, but the FoMs eventually start to drop when removing more bins.

When fixing bias (blue dotted line) the absolute constraints greatly improve, as expected from breaking degeneracies. $\fom$ shows substantially relative lower oscillations (bottom left panel), which means that part of the improvement came form better measuring bias, while $\fomg$ shows greater oscillations when bias is fixed. $\fomc$ combines these effects and improves a factor 1.2 at the peak when fixing bias (10\% lower than the fiducial). When removing RSD (purple dot-dashed), the constraints for $\gamma$ reduce considerably and look flat in the absolute values (center right panel). $\fomg$ shows higher oscillations (center left panel), but now these come from measuring $\gamma$ directly from the growth rate in front of the power spectrum, and not from RSD. In $\fom$ the constraints are worse, but the relative gains are very similar.

\xfigurefull{}{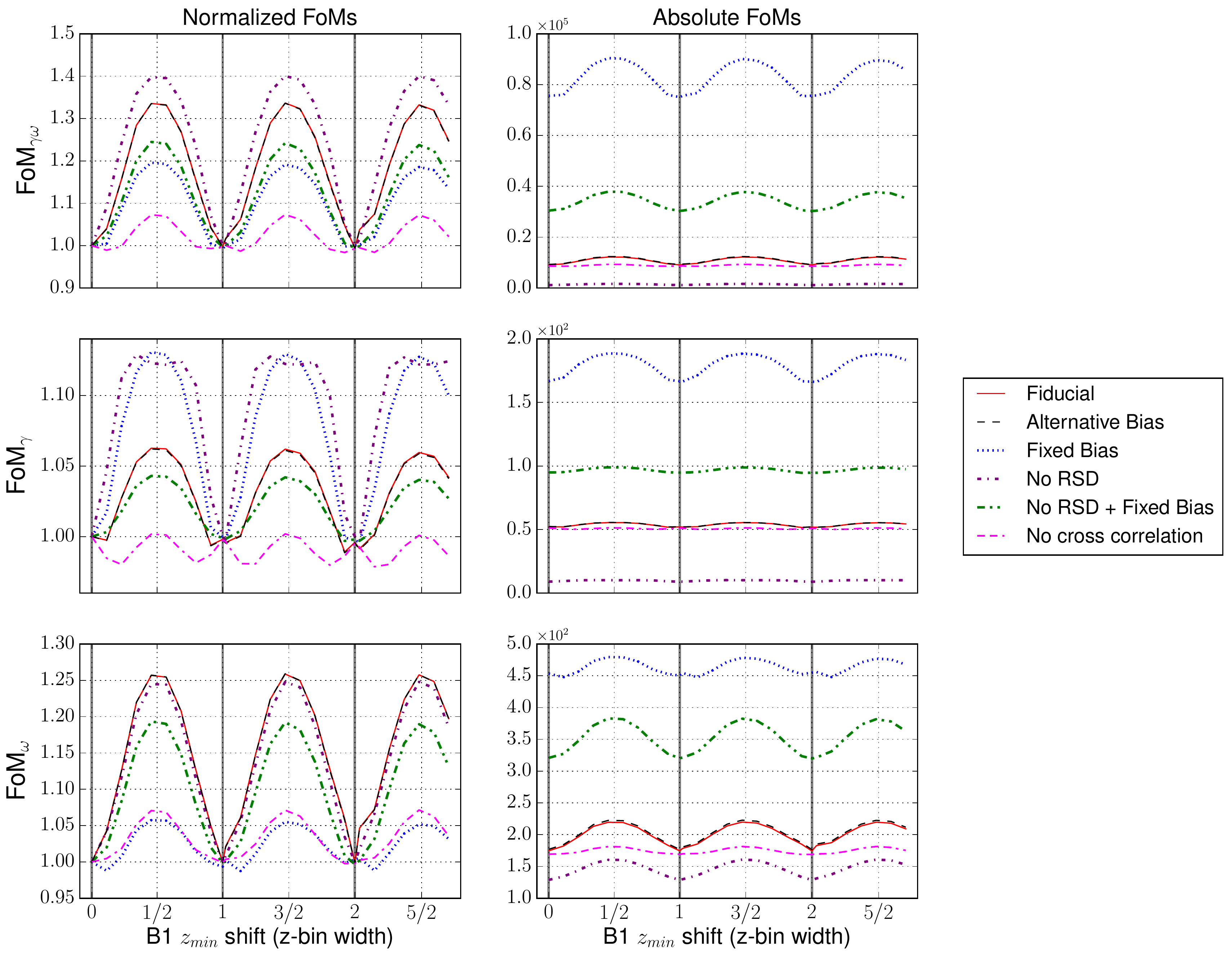}
{Effect of having partly overlapping redshift bins when combining two spectroscopic surveys (B1xB2). The start $z_{min}$ of the B1 redshift range determines the overlap between the redshift bins of both populations. The x-axis shows the B1 $z_{min}$ shift in z-bin width units, $\Delta z$. The panels on the left show the FoM normalized to the fully overlapping bins value (i.e. B1 $z_{min}=0.1$ or shift $=0$) for $\fomc$, $\fomg$ and $\fom$, while the panels on the right show the absolute values. The black (dashed) line uses 4 redshift pivot points $z_i\in[0.25, \,0.43, \,0.66,\, 1.0]$ to parametrize bias instead of the fiducial 1 parameter per redshift bin and population. The pink dash-dash-dot line does not include cross correlations between B1 and B2. The blue dotted line is the fixed bias case, the purple dot-dashed line corresponds to removing RSD, and the green dash-dot-dot line combines fixed bias and no RSD. The grey vertical lines show the fiducial (B2) redshift bin edges.}
{fig:fom_zmin_abs}

Fig. \ref{fig:fom_zmin_width} shows the impact of the redshift bin width on the oscillations in $\fomc$. We parametrize the bin width as $\Delta z = w(1+z)$. The lines correspond to: $w=0.01$ (red solid, fiducial value), $w=0.0075$  (blue dashed) and $w=0.0125$ (green dotted). All lines are normalized to their respective values at B1 $z_{min}=0.1$. It shows that redshift bin width has an important impact on the relative gains. For the thinner binning the relative improvement is only of a factor 1.2, while for the thick binning is $\sim$1.5 (the fiducial is 1.33). This shows that if the binning is narrower the relative gain is lower as the radial resolution is better, but recall that the maximum resolution is limited by only using linear scales ($k_{max}$).

\xfigure{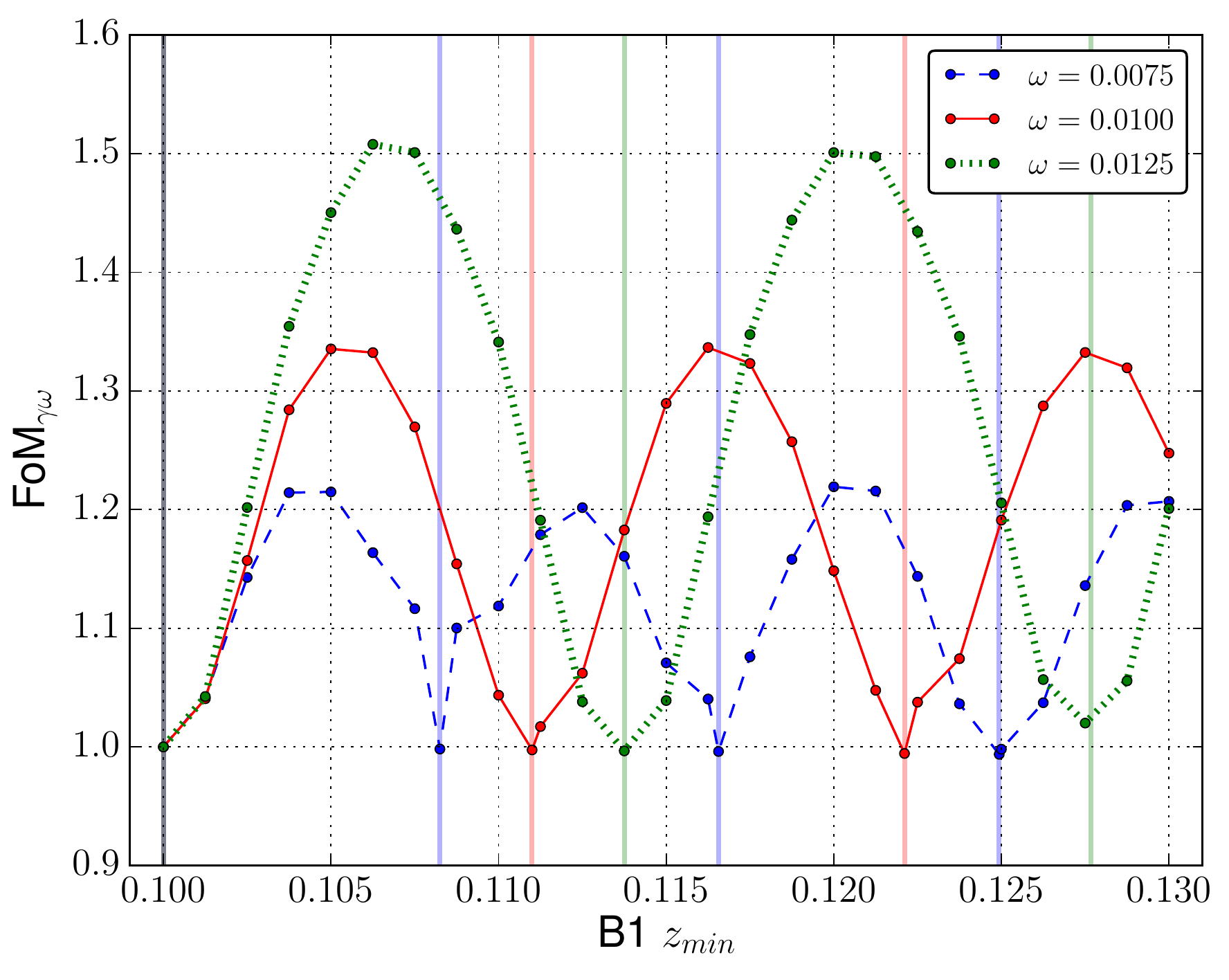}
{Amplitude of the oscillations when having partly overlapping bins between two populations, for different bin widths. For $\Delta z=w(1+z)$, the lines correspond to: red solid, $w=0.01$ (fiducial value); blue dashed, $w=0.0075$; green dotted, $w=0.0125$. All lines are normalized to their respective values at B1 $z_{min}=0.1$. The vertical lines show the position of the bin edges for each case.}
{fig:fom_zmin_width}

Fig. \ref{fig:fom_bamp_zmin} shows B1xB2 as function of the relative bias amplitude for different B1 $z_{min}$ shift values, which shows the impact of partially overlapping bins from having full overlap (B1 $z_{min}$ shift $= 0.0 \,\Delta z$, red solid line) to almost half overlap (B1 $z_{min}$ shift $=0.45\,\Delta z$, blue dashed). In $\fomg$ (top right panel) increasing the partial overlap has several effects. When bias amplitudes are similar there is more gain from partial overlap, while when the bias amplitude grows this gain decreases until the point that shifting bins leads to worse constraints. Also, for half overlapping bins $\fomg$ flattens for $\alpha>1$. For $\fom$ and $\fomdetf$ there is always gain from partial overlap. The different lines are closer for lower $\alpha$ where the gain is minimum, which increases until $\alpha=1$. From that point the lines are quite parallel. When there is full overlap we have the minimum at $\alpha=1$ (same bias case) and the FoMs increase with the bias difference, but when there is half overlap between the redshift bins of both populations B1xB2 behaves like a single tracer, in the sense that $\fomg$ decreases with bias while $\fom$, $\fomdetf$ increase with bias (see Fig. \ref{fig:fom_redshift_bamp}). On the other hand, $\fomc$ combines the effects from $\fomg$ and $\fom$ and keeps the minimum, increasing the FoM for higher partial overlap, meaning the gain is higher when bias is similar. 

\xfigurefull{0.75}{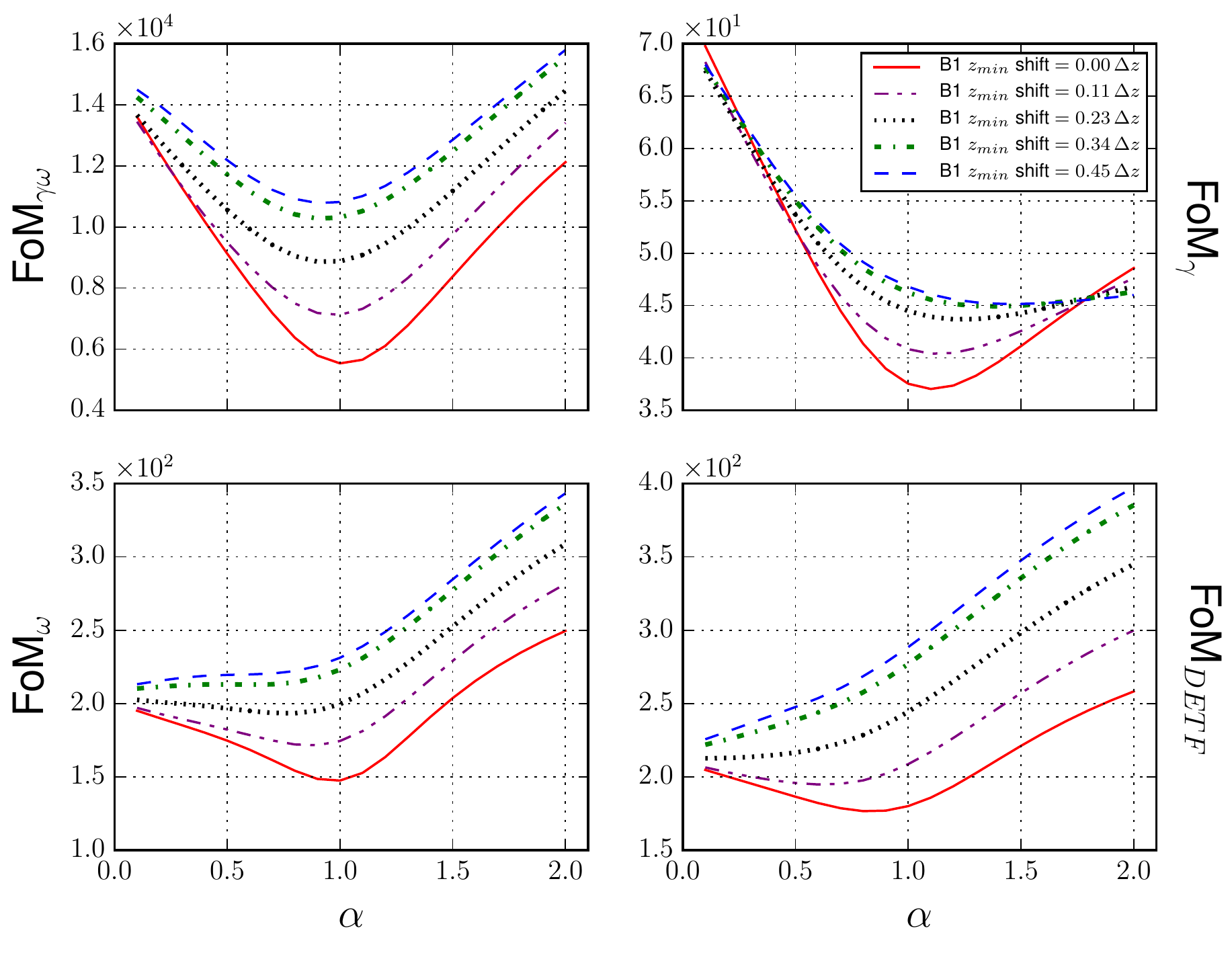}
{Effect of partially overlapping bins in B1xB2 as function of the relative bias amplitude between both populations ($\alpha$). The lines correspond to B1 $z_{min}=$ 0.1 (red solid), 0.10125 (green dash-dot), 0.1025 (black dotted), 0.10375 (purple dash-dot-dot), 0.105 (blue dashed).}
{fig:fom_bamp_zmin}

\subsection{Radial resolution} \label{subsec:Radialcorrelations}  

In this subsection we study the impact of increasing the number of spectroscopic redshift bins. In the fiducial forecast we use spectroscopic surveys with 71 narrow redshift bins, such that at each bin we mainly account for transverse modes from angular spectra, while the radial information (modes) is contained in the cross correlations between redshift bins. This tomography study can approximately recover the full 3D clustering information when the comoving redshift bin separation, $\Delta r = c\Delta z / H(z)$, corresponds approximately to the minimum linear 3D scale $\lambda_{min}^{3D} = \frac{2\pi}{k_{max}}$, (\citealt{2012MNRAS.427.1891A}). As we are limited by the linear regime, including more bins would eventually lead to include nonlinear modes, which would require modeling the nonlinear angular power spectrum. The bin width $\Delta z=w(1+z)$ is set by the number of redshift bins,

\begin{equation}
w = \sqrt[\leftroot{-2}\uproot{9}N_z]{\frac{1+z_{max}}{1+z_0}}-1,
\end{equation}

\medskip

\noindent which divide the interval $[z_0, z_{max}]$ into $N_z$ redshift bins (see \citealt{2015arXiv150203972E}).

Fig. \ref{fig:fom_nbins} shows how the constraints improve when increasing the number of redshift bins for $\fomc$. The lines show B1 (blue dotted), B2 (green dot-dash), B1xB2 increasing both B1 and B2 redshift bins (red solid), and B1xB2 keeping fixed B1 number of redshift bins to 71 (black dashed). Both single populations  and combined surveys improve when increasing the number of bins. There are several effects when we increase the number of redshift bins $N_z$ (see \citealt{2012MNRAS.427.1891A}). As the redshift bin width corresponds approximately with the minimum scale, increasing the number of redshift bins in the same redshift range allows for probing smaller scales, which gives more independent modes and improve constraints. The signal to noise at each bin remains nearly constant as the auto power spectrum has higher clustering from having more close pairs while the density per bin is lower, increasing both signal and shot noise. Therefore, increasing $N_z$ increase the number of transverse modes without lowering their signal to noise, which improve constraints. Eventually, the redshift bins will become correlated and this gain will saturate (see \citealt{2015arXiv150203972E}). In addition, when increasing $N_z$ we also add more cross-correlations between redshift bins, which is very important as RSD depends on the relation between radial and transverse modes ($\mu= k_\|/k$). 

\medskip

\xfigure{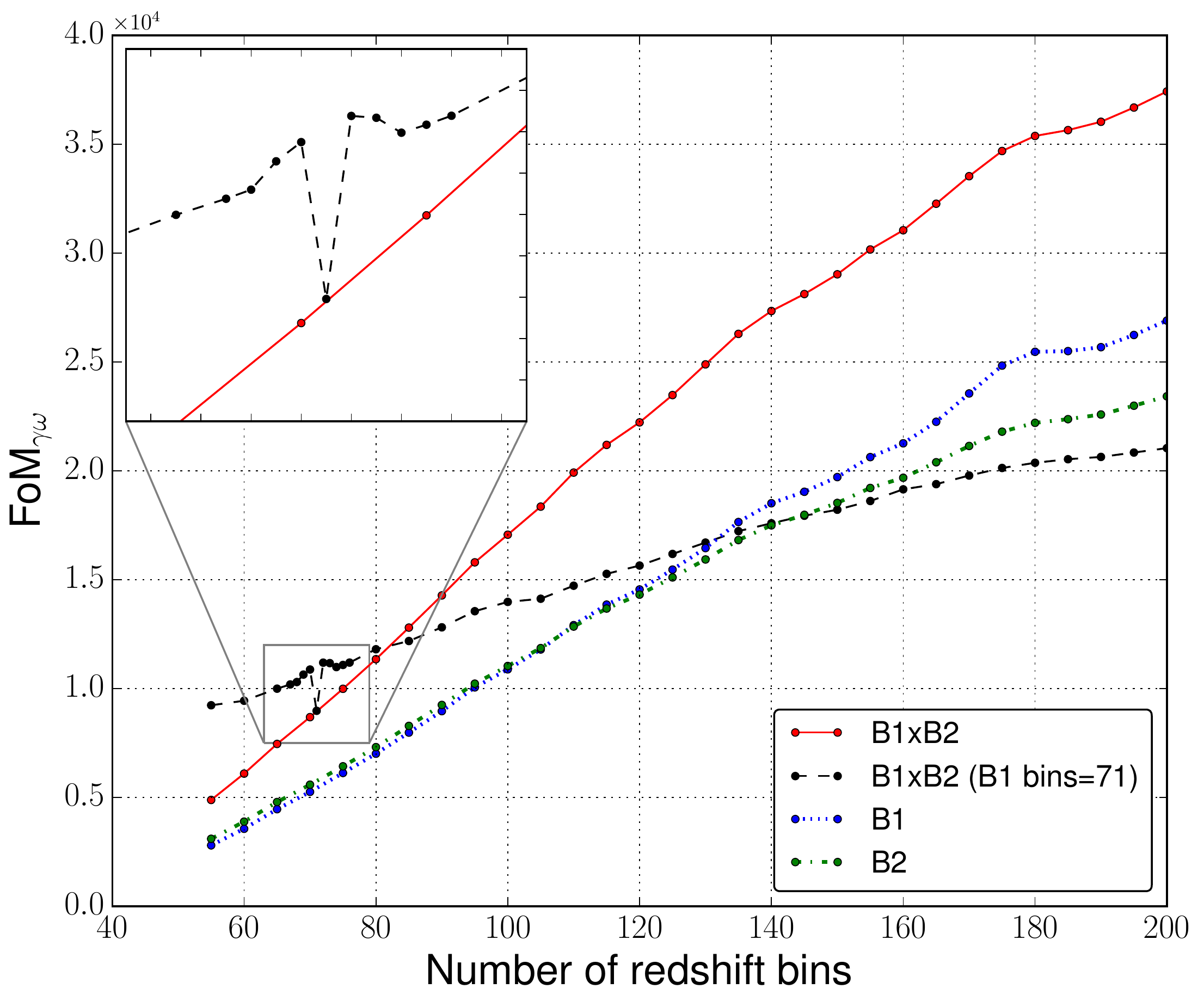}
{Effect of increasing the number of spectroscopic redshift bins in $\fomc$. The lines correspond to: B1xB2 (red solid), B1xB2 with B1 bins fixed to 71 (black dashed), B1 (blue dotted) and B2 (green dot-dashed).}
{fig:fom_nbins}

In Fig. \ref{fig:fom_nbins} we also study the effect of fixing the bins of one of the populations in the overlapping survey. The black dashed line refers to B1xB2 with the B1 redshift bins fixed to 71, while the number of B2 bins vary. This results in a flatter improvement than when increasing the bins for both populations as we are adding less redshift bins. An interesting behavior happens when B1 and B2 have similar (but different) number of redshift bins (zoomed region). As we have previously discussed, having two populations with partly overlapping bins improve the constraints. $\fomc$ improves by a factor 1.2 (1.25) when using one less (more) redshift bins in B2. This gain is equivalent to be using $\sim80$ redshift bins for both populations instead of the fiducial 71.

\xfigure{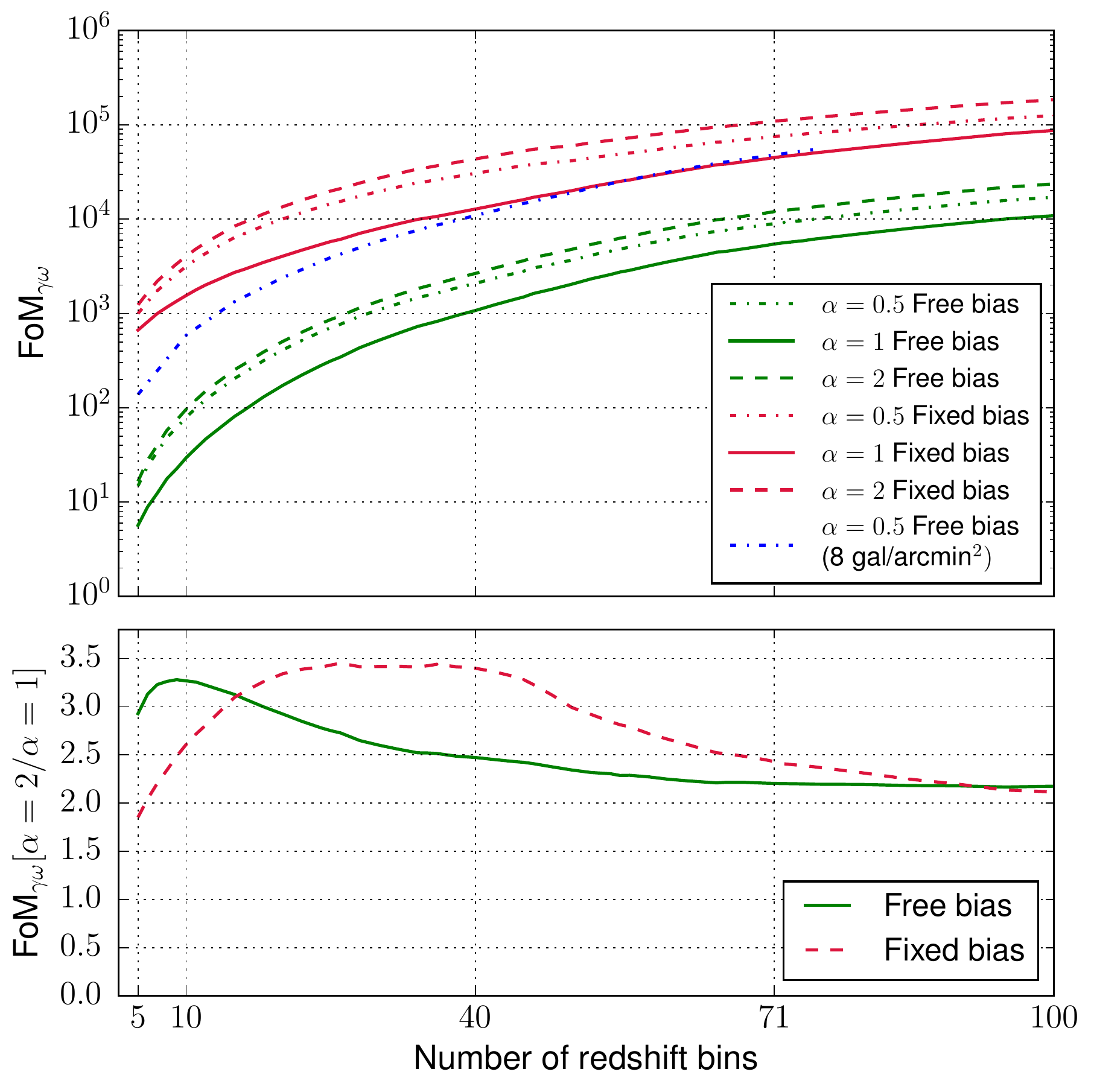}
{Effect of reducing the radial information by using less number of redshift bins, keeping the fiducial density (0.4 gal/arcmin$^2$). A binning of $5\sim10$ bins corresponds to typical photometric surveys, a binning of 40 correspond to narrow photometric surveys and 71 is the fiducial binning for spectroscopic surveys in this work. Top panel shows $\fomc$ for $\alpha=0.5,1.0, 2.0$ for free and fixed bias. Bottom panel shows the ratio $\fomc[\alpha=2/\alpha=1]$. Free bias with $\alpha=0.5$ and density $8$ gal/arcmin$^2$ is shown for comparison as a typical photometric survey density.}
{fig:fom_surveys}

\subsubsection{Sample variance cancellation in photometric surveys} \label{subsec:svcphotoz}

Using thicker redshift bins we can model the loss of radial information and study the impact of sample variance cancellation in photometric surveys. $\text{Fig. }\ref{fig:fom_surveys}$ shows $\fomc$ with ($\alpha=0.5,\,2.0$) and without ($\alpha=1.0$) cancellation for a smaller number of redshift bins (recall that with $\alpha=1$ cancellation is no longer possible). We show the effect for typical photometric surveys such as DES or EUCLID ($5\sim10$ bins), for narrow photometric surveys like PAU ($\sim40$ bins) and compare it to spectroscopic-like surveys (fiducial 71 bins). The top panel shows that $\fomc$ improves strongly when increasing the radial resolution, specially in the free bias case where it improves 3 orders of magnitude from 5 to 100 bins as opposed to 2 orders in the fixed bias case. Radial cross-correlations are more important in a free bias forecast, which leads to more radial dependence. Bottom panel shows the sample variance cancellation effect in photometric surveys in the ratio between $\alpha=2.0$ and $\alpha=1.0$. We find that the contribution from sample variance cancellation has a larger or at least similar effect in photometric surveys for both free and fixed bias. Note how density (e.g. compare to Fig. \ref{fig:foms_densityX}) is as important as redshift accuracy in that it can also change $\fomc$ by 3 orders of magnitude.

Table \ref{surveystable} shows there is a strong density dependence in the ratio, specially in surveys with less number of redshift bins. While for spectroscopic surveys (e.g. 71 bins and 0.4 gal/arcmin$^2$) the ratios are 2.2 and 2.4 for free and fixed bias, for photo-z surveys (e.g. 5 bins and 8.0 gal/arcmin$^2$) the ratios become 19.5 and 5.32. Surveys with a low number of redshift bins are more dominated by radial auto-correlations, which are affected by shot-noise, so they become more density dependent, as shown in Table \ref{surveystable}. For a survey density of 0.4 gal/arcmin$^2$ auto-correlations contribute up to $72\%$ for a survey with 10 redshift bins with free bias, while only $29\%$ and $10\%$ for 40 and 71 bins. This high density dependence explains the turnover in the free bias ratio for low number of bins in Fig. \ref{fig:fom_surveys}, where shot-noise limits the gains from sample variance cancellations, and also produces very large gains for higher densities as shown in Table \ref{surveystable}. 

Concerning absolute FoMs for photo-z and spec-z surveys, the loss of radial resolution from photometric uncertainties is in some part compensated by a gain in density in photometric surveys. The combined figures of merit $\fomc$ for a survey without split take the values of 7.19 and 5453 for photo-z (5 bins and 8.0 gal/arcmin$^2$) and spec-z (71 bins and 0.4 gal/arcmin$^2$) surveys. Using the fiducial configuration ($\alpha=0.5$) with sample variance cancellation leads to a $\fomc$ of 138 and 8979, reducing the ratio between the FoM between both surveys by a factor $\sim12$ (compare blue and green lines in Fig. \ref{fig:fom_surveys}).

\begin{table}\tiny \centering
  \begin{tabular}{ c c c c c c  }
    \hline
    Bias & \hspace{-5mm}\begin{tabular}[c]{@{}c@{}}Density\\(gal/arcmin$^2$)\end{tabular}& 5 bins & 10 bins & 40 bins & 71 bins \\ \hline
	& 0.4 & 2.93 & 3.27 & 2.47 & 2.20 \\
    Free  & 8.0 & 19.5 & 16.2 & 8.81 & 6.94\\
    & 80.0 & 138 & 93.4 & 66.1 & 38.4 \\ \hline
    & 0.4 & 1.85 & 2.61 & 3.40 & 2.43 \\
    Fixed & 8.0 & 5.32 & 7.10 & 8.80 & 5.29 \\
    & 80.0 & 13.4 & 15.2 & 29.3 & 21.3 \\ \hline
   \end{tabular}
\caption{Ratio between $\fomc[\alpha=2/\alpha=1]$, which shows the relative contribution of sample variance cancellation (see the text). On the columns there are 4 different binnings (5, 10, 40 and 71 redshift bins), while on the rows there are 3 different total survey densities for free and fixed bias. }
\label{surveystable}
\end{table}

%% file: conclusions_clean.tex
\section{Conclusions} \label{sec:conclusion} 

In this paper we have estimated dark energy ($\omega_0$, $\omega_a$) and growth rate ($\gamma$) constraints of multiple tracers in spectroscopic surveys using the Fisher matrix formalism. In the fiducial forecast we use galaxy clustering from 2D angular correlations in 71 narrow redshift bins which include baryon acoustic oscillation (BAO) in the linear Eisenstein-Hu power spectrum (\citealt{1998ApJ...496..605E}) and the linear Kaiser effect (\citealt{1987MNRAS.227....1K}) to account for redshift space distortions (RSD). To compress the information of our constraining power in one number we define four Figures of Merit: $\fomc$, $\fomg$, $\fom$ and $\fomdetf$ ($\S$\ref{subsec:fom}). Details of the modelling and fiducial forecast assumptions can be found in section \ref{sec:theobkg}.

Section \ref{sec:samplevar} studied how multiple tracers in the same region of the sky can break degeneracies and improve the constraints on dark energy and growth rate. We split one spectroscopic survey into two populations (named B1 and B2) by some galaxy property such that there is a relative bias amplitude ($\alpha$) between both populations. This allows for multi-tracing the same underlying matter distribution, which cancels the random nature (sample variance) of fluctuations in redshift space. Using redshift space distortions and two populations as multiple tracers we quickly improve the constraints when increasing the bias difference between the tracers, with no or very little improvement with no bias difference (Fig. \ref{fig:fom_redshift_bamp_densgal_01}). Fixing the bias ratio to $\alpha=0.5$, we increased the galaxy density and showed that B1xB2 FoMs outperform the single tracers by beating the sampling variance limit in the noiseless limit (Fig. \ref{fig:foms_densityX}). We showed that this improvement comes from sample variance cancellation and not from additional cross-correlations between B1 and B2, which contribute less than 2\%. Also, Fig. \ref{fig:foms_xbias} showed that not all improvement was coming from measuring bias. 

In section \ref{sec:clfhm} we have set up an r-band limited survey with a CLF and HM to model galaxy bias. We have split the survey into two subsamples cutting in halo mass and r-band magnitude, computing the galaxy bias and galaxy density of each subsample in redshift. In this way, we account for the relation between galaxy bias and galaxy density. We showed that for a cut in halo mass we can improve $\fomc$ up to a factor 1.3  as compared to doing no split. Splitting in r-band magnitude lead to a factor 1.05 improvement as magnitude scatter halo mass which reduced the bias difference. When increasing the total density of the survey we found huge improvements for both split methods, giving a factor 9.6 and 3.0 in $\fomc$ for halo mass and r-band cut, respectively.

In section \ref{sec:Partlyoverlapping} we have studied the effect of having partially overlapping redshift bins in a multiple tracer survey. We have shown   that as a result of the overlap the FoMs improve (Fig. \ref{fig:fom_zmin_abs}), having a peak when the redshift bins are shifted half of the bin width. At the peak the FoMs improve a factor 1.33 for $\fomc$, 1.06 for $\fomg$, 1.26 for $\fom$ and 1.33 for $\fomdetf$. We have shown that the gain is not artificially produced by the particular bias parameterization, but it is rather coming from the cross-correlations between B1 and B2. We have also shown that the gain is a factor $\sim1.5$ in $\fomc$ when using a 25\% thicker binning, while a factor $\sim1.2$ when using a 25\% thinner binning (Fig. \ref{fig:fom_zmin_width}). Fig. \ref{fig:fom_bamp_zmin} showed how FoMs improve from partly overlapping bins for different $\alpha$ values, indicating that there is more improvement when relative bias difference is small, and low or no improvement when there is more bias difference between populations. In Fig. \ref{fig:fom_nbins} we showed how FoMs improve when increasing the number of redshift bins. When using one more or less redshift bin in one of the populations, the improvement in $\fomc$ is equivalent to having 80 redshift bins in both populations compared to the 71 from the fiducial forecast. In  section \ref{subsec:svcphotoz} we find that the multi-tracer gains are larger for photometric samples, specially when we increase the density. In fact, having a larger density can compensate the loss in resolution (see Fig. \ref{fig:fom_surveys}).

In $\S$\ref{subsubsec:reds} to \ref{subsubsec:rsdbao} we have studied the impact of RSD and BAO effects in the FoMs. In Fig. \ref{fig:fom_redshift_bamp} we show the improvement from B1xB2 for different bias difference in the four FoMs. The constraints quickly improve in all FoMs, up to a factor 4 in $\fomc$. In real space (no RSD, Fig. \ref{fig:fom_redshift_bamp_real}) sample variance cancellations are no longer possible and the improvements are less than a factor 1.5 in $\fomc$, having almost no gain in $\fomg$ (less than 2\%). In Fig. \ref{fig:fom_redshift_bamp_rsd_bao} we have shown that RSD is very important in $\gamma$ constraints as it breaks degeneracy between galaxy bias and $f(z)$. For $\fomg$ the B1xB2 ratio RSD/No RSD is $4\sim9$, while the ratio BAO/No BAO is lower than $1.08$. For dark energy constraints it happens the opposite, in $\fom$ the BAO/No BAO rate is $1.4\sim 1.5$ while RSD/No RSD is $\sim1.0$ when bias amplitudes are similar, but up to $\sim 1.5$ when there is more bias difference. The $\omega$ constraints depend more on the shape of the power spectrum and then are enhanced more from including BAO wiggles.

In $\S$\ref{subsubsec:xcosmo} we have fixed cosmological parameters to study possible degeneracies. We find that multiple tracers help breaking degeneracies in the dark energy constraints, specially at high densities. In $\S$\ref{appB} we studied the impact of a different bias evolution slope in redshift (Fig. \ref{fig:fom_zrange_biaslope}). It shows a dependence for the B1 population, but the B1xB2 forecast is insensitive to the bias slope.

The trends found in Section 3 are in good agreement with a number of studies such as \cite{2008.0810.0323}, \cite{2009MNRAS.397.1348W}, \cite{2010MNRAS.407..772G}, \cite{2011MNRAS.416.3009B}. Direct quantitative comparison is hard to make due to very different survey configurations, along with different modeling and observables. \cite{2010MNRAS.407..772G} suggested that cuts other than halo mass such as peak-height $\nu$ might be more competitive for a dark matter haloes split, as they find a $\sim10\%$ improvement at low redshift with a cut in halo mass. In section \ref{sec:clfhm} we have stressed that these gains are highly density dependent and that the split for galaxies can be optimized by looking at the bias-density relation of the tracers. 

In this analysis we have assumed a number of idealizations, such as linear theory, deterministic bias, no stochasticity between tracers nor nonlinearities. Several studies indicate that linear theory (\citealt{1987MNRAS.227....1K}) start to break down at scales as large as $k_{max} > 0.02\, h\text{Mpc}^{-1}$  (\citealt{2011ApJ...726....5O}, \citealt{2012MNRAS.427.2420B}), specially at low mass haloes, and that scale dependence in $\beta$ varies between tracers. As shown in \cite{2010MNRAS.407..772G} even small amounts of nonlinearity can degrade your FoM down to 50\%, which emphasize the need for more realistic models for galaxies and nonlinear RSD (e.g. \citealt{2011MNRAS.417.1913R}, \citealt{2015PhRvD..92j3516O}). Lately, a number of techniques have been developed to reduce shot-noise and stochasticity under the Poisson level by optimally weighting the tracers (see \citealt{2009PhRvL.103i1303S}, \citealt{2010PhRvD..82d3515H}, \citealt{2011MNRAS.412..995C}, \citealt{2016MNRAS.tmp.1283P}) which become of most interest in combination with multi-tracer surveys and could further improve the FoMs.

In summary, our results suggest that we can improve FoM significantly and break degeneracies in cosmological inference if we  split the samples by a density ratio of $\bar{n}_{1}/\bar{n}_{2} \sim 7$ using apparent magnitude as ranking or $\bar{n}_{1}/\bar{n}_{2} \sim 30$ using a mass halo proxy ranking (e.g., richness). Using another proxy for bias (such as local density, see \citealt{2015arXiv151001692P}, or color) to split the sample, will give similar benefits. Splitting volume limited samples does not provide significant improvements. Our analysis also shows that when doing angular clustering tomography is optimal to use overlapping bins for cross-correlation. These finding can be applied to future redshift surveys such as DESI or Euclid and will also work for photometric samples (see Fig. \ref{fig:fom_surveys}), such as DES and LSST and its cross-correlations (as shown in \citealt{2015arXiv150203972E}). We also show large improvement on the FoM with increasing galaxy density. This can be used as a trade-off to compensate a possible loss of radial resolution when using high resolution photometric redshifts (\citealt{2014MNRAS.442...92M}) instead of spectroscopic redshifts.

%% file: acknowledgements_clean.tex
\section*{acknowledgements} \label{sec:acknowledgements} 
A.A. and E.G. acknowledge support from the Spanish Ministerio de Ciencia e Innovacion (MICINN), project AYA2012-39559 and AYA2015-71825,  and research project 2014 SGR 1378 from Generalitat de Catalunya. M.E. acknowledge support from the European Research Council under FP7 grant number 279396. Calculations in this paper have been done using  the Port d’Informacio Cientifica (PIC, www.pic.es) computation infrastructure. We thank Jorge Carretero for help at PIC.

%% file: appendix_A_clean.tex
\appendix

\section{RSD and BAO effects in sample variance canellation and degeneracies}  \label{appA} 

\subsection{Redshift space}  \label{subsubsec:reds} 

In Fig. \ref{fig:fom_redshift_bamp} we present our results for the four FoMs. The blue (dotted) lines corresponds to the same sky case (B1xB2), the green (dashed) lines to B2 and the red (solid) lines to B1, with all lines being normalized to the B1 lines. Recall from Eqs. \ref{b1}-\ref{b2} that a change in $\alpha$ modifies the B2 bias but leaves the B1 bias equal to the fiducial. Focusing first on the B1xB2 lines, the four FoMs show an improvement compared to the single tracer cases, as expected, due to sample variance cancellations. We can see how the improvement for B1xB2 increases with the bias difference and is minimum when B1 and B2 have the same bias, as then we cannot cancel sample variance. $\fomc$ shows the biggest gains (factor of 4 and 3 at $\alpha\simeq 0.1$ and $\alpha\simeq 2$) as it combines the gains from $\fomg$ (factor of 2) and $\fom$ (factor of 2).
Part of the gain comes from B1xB2 doubling the density of the single tracers, as we have merged both single tracers and it reduces shot-noise. To see the effect of splitting B1 into B1xB2 to keep the shot-noise level see Fig. \ref{fig:foms_densityX}.

The B2 lines show the impact that bias amplitude has on the single survey case. $\fomg$ shows that $\gamma$ constraints improve for lower bias. This is in the line of \cite{2014MNRAS.445.2825A}, where the authors find similar results for the dependence of $\gamma$ constraints on bias for a photometric survey. This is because lower bias gives larger relative importance to RSD, which turns into a better measurement of $f(z)$ and thus $\gamma$.  On the contrary, $\omega$ constraints improve with larger bias, as shown in $\fom$ and $\fomdetf$. This is due larger bias increasing the signal of correlations and thus reducing the relative impact of shot noise (see Fig. \ref{fig:fom_zrange_biaslope} bottom right panel for similar trends).

\xfigure{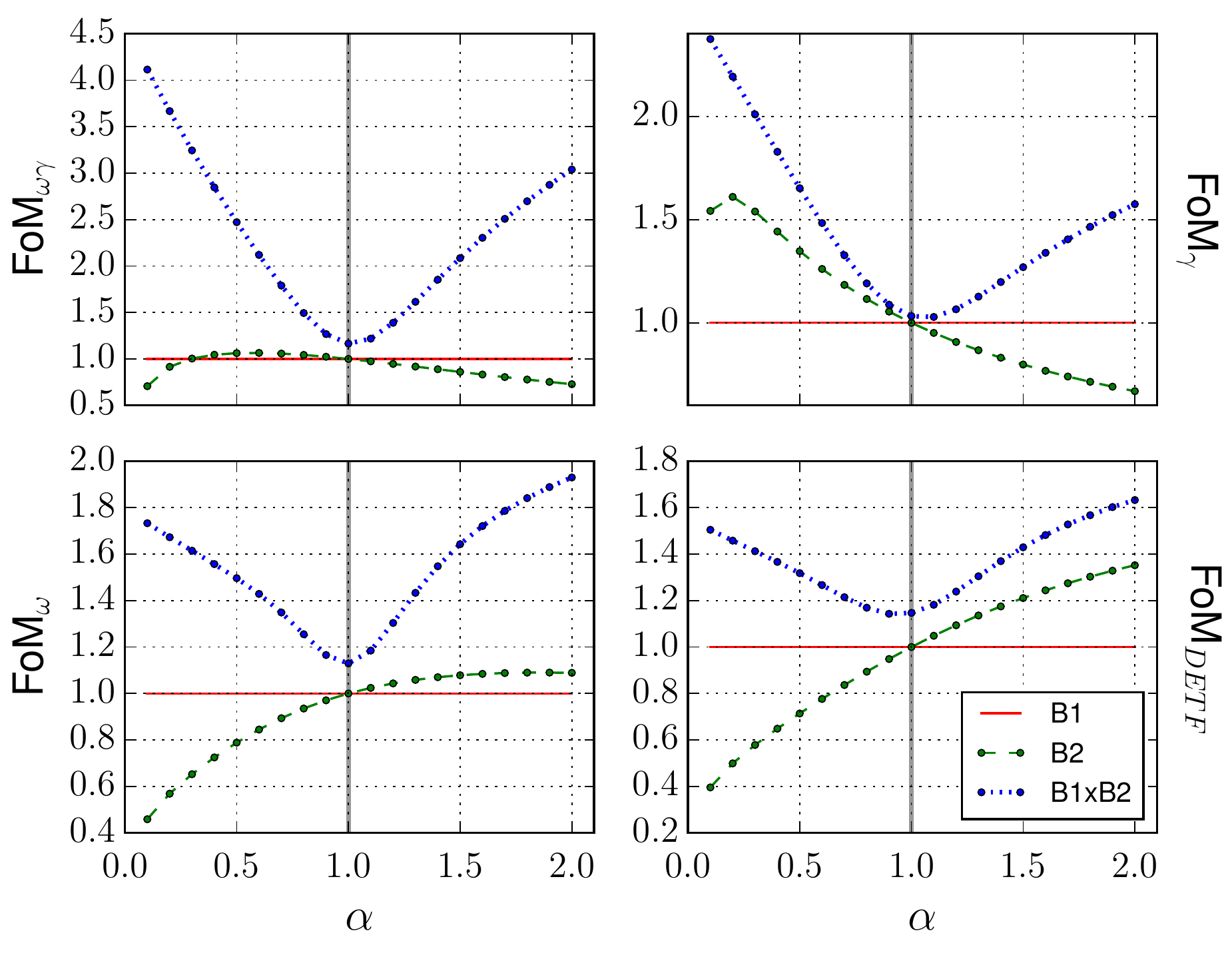}
{FoM dependence on the relative bias amplitude ($\alpha$) in redshift space. The blue (dotted) line correspond to the same sky case (B1xB2), the green (dashed) line to B2 and the red (solid) line to B1. All lines are normalized to the B1 FoM value (which is constant). The vertical lines indicate the same bias amplitude case ($\alpha=1$).
}
{fig:fom_redshift_bamp}

\subsection{Real space}  \label{subsubsec:reals}  

\xfigure{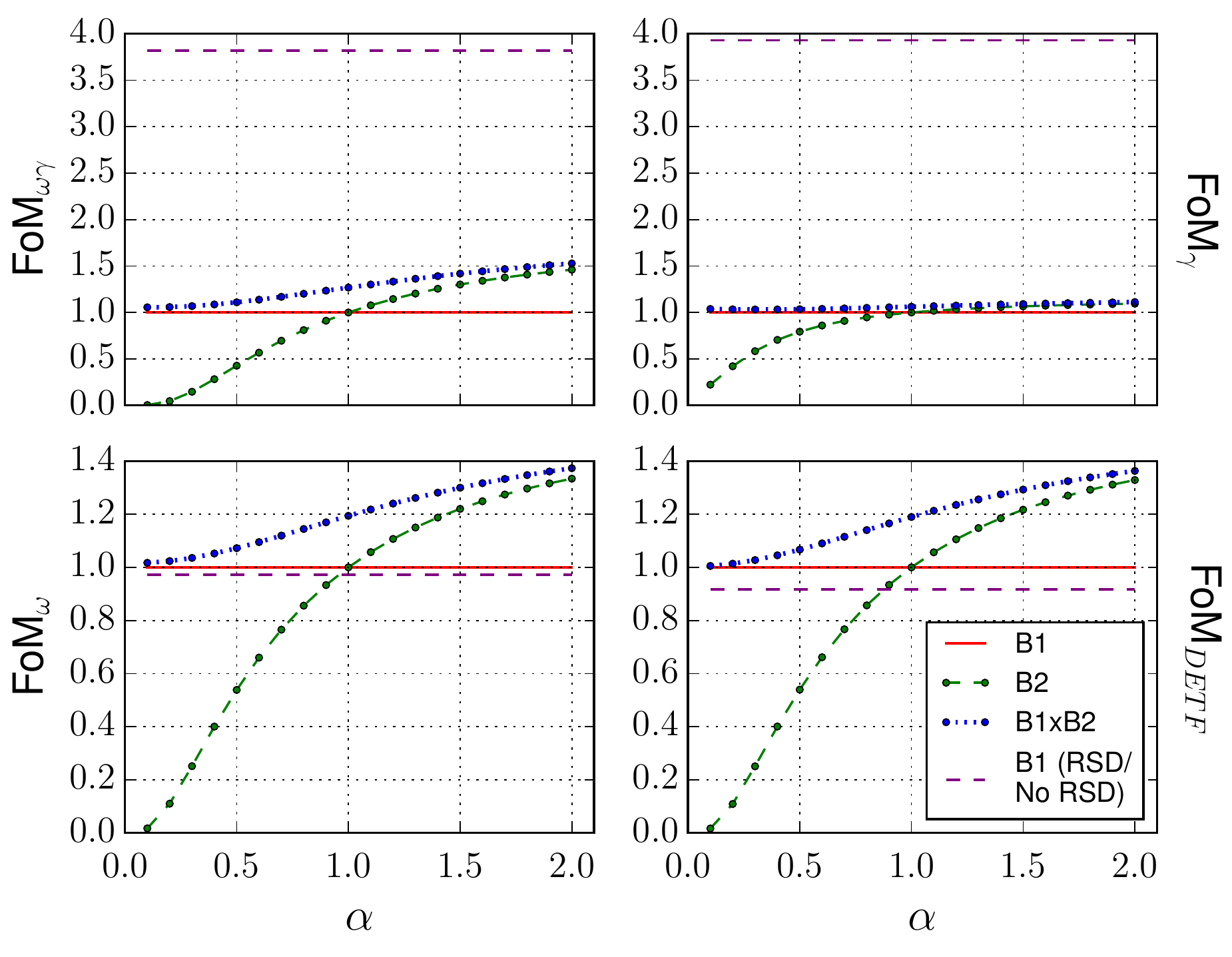}
{FoM dependence on the relative bias amplitude ($\alpha$) in real space. The blue (dotted) line correspond to the same sky case (B1xB2), the green (dashed with dots) line to B2 and the red (solid) line to B1. All lines are normalized to the B1 FoM value (which is constant). The purple (dashed) line shows the ratio RSD/No RSD for the fiducial bias (B1).}
{fig:fom_redshift_bamp_real}

In Fig. \ref{fig:fom_redshift_bamp_real} we have removed redshift space distortions (RSD). It includes a purple (dashed, flat) line which shows the ratio RSD/No RSD for B1. Looking at the B1xB2 line we see how all FoMs now grow with bias ratio, while the characteristic minimum and big improvements from sample variance cancellation have vanished (factor of 1.5 for $\fomc$ now, as opposed to 4 with RSD).

In real space the density perturbation equation is Eq.\ref{galaxybias}, which does not have specific angular dependence and does not allow to cancel sampling variance anymore. The auto correlations for a redshift bin $i$ will then be proportional to $b_i^2$, while the cross correlations between a bin $i$ and a bin $j$ is proportional to $b_i b_j$. Therefore, signal will increase as bias does. Moreover,

\begin{equation}
\begin{split}
\text{when } \alpha\rightarrow 0 \text{ then } \text{FoM[B1xB2] $\simeq$ FoM[B1]}, \\
\text{when } \alpha\gg 1 \text{ then } \text{FoM[B1xB2] $ \simeq$ FoM[B2]}.
\end{split}
\end{equation}

This is a result of B1 correlations dominating over B2 ones at $\alpha\simeq 0$, and viceversa for $\alpha\gg 1$. Also, the B2 FoMs have a steeper slope than B1xB2, as we are only changing the B2 bias, which only affects a subset of correlations on B1xB2. While $\fom$ improves up to a 40\% at $\alpha=2$, the gain in $\fomg$ is marginal, and part of it comes from B1xB2 having more density and reducing shot-noise.

The purple (dashed flat) line shows the ratio between the normalizations when computing the correlations in redshift space and in real space. This shows the overall impact of removing RSD in each FoM. As can be seen in the top-right panel, $\fomg$ is about 4 times lower, as RSD are crucial for breaking degeneracies between $f(z)$ and $b(z)$, which affect the $\gamma$ constraints. For the $\omega$ constraints, the normalizations are similar. However, as we have discussed, including RSD is still very important as it allows to cancel sampling variance and improve the $\omega$ constraints.

\subsection{Relative impact of RSD and BAO }  \label{subsubsec:rsdbao} 

In Fig. \ref{fig:fom_redshift_bamp_rsd_bao} we explore the effect that removing RSD  and BAO has on the absolute value of the FoMs, for different relative bias amplitudes ($\alpha$). We show the ratios `RSD/No RSD' (left panels) and `BAO/No BAO' (right panels), where `RSD' and `BAO' refers to the fiducial forecast that includes both RSD and BAO effects. The left/right panels correspond to removing RSD/BAO for B1xB2 (blue solid), B2 (green dashed) and B1 (red dotted). 

\xfigure{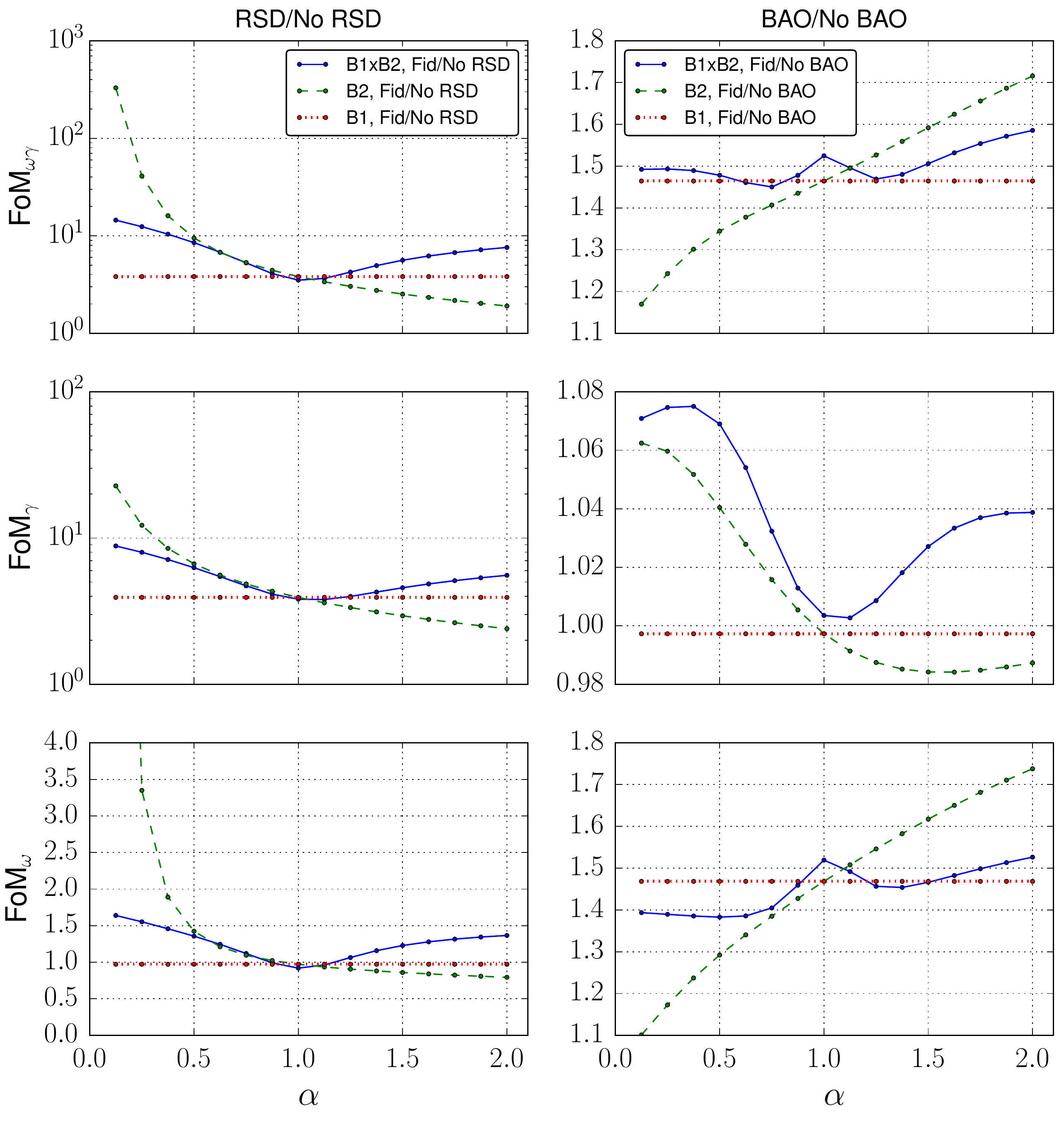}
{Impact of removing BAO and RSD effects on the constraints, for different relative bias amplitudes ($\alpha$). The left/right panels correspond to removing the RSD/BAO effect for B1xB2 (blue solid), B2 (green dashed) and B1 (red dotted). All the lines show the ratio between the Fiducial forecast where RSD and BAO effects are included and when removing one of the effects. There is no further normalization.}
{fig:fom_redshift_bamp_rsd_bao}

The trends in $\fomg$ (center panels) show how RSD has a strong relative importance in $\gamma$ constraints, while BAO in comparison has very little impact. On the other hand, $\fom$ (bottom panels) shows how BAO is more important to measure $\omega$ than RSD. This happens because measuring $\gamma$ depends more on determining the amplitude of the power spectrum $P(k)$, which is enhanced by RSD as it breaks degeneracies between bias and the growth factor, while $\omega(z)$ measurements comes more from measuring its shape and thus including BAO wiggles measurements improves it. In addition, for $\fom$, B1xB2 ( solid) shows a peak at $\alpha=1$ in BAO/No BAO ratio, because the BAO contribution is relatively higher when RSD is less important. On the other hand, for $\fomg$ the BAO contribution in B1xB2  gets enhanced with a better RSD signal. The impact of RSD in B1xB2 is higher for more bias difference in all FoMs, as previously discussed (see $\S$\ref{subsubsec:reds}), due to sample variance cancellations.

For a single tracer (see B2 in Fig \ref{fig:fom_redshift_bamp_rsd_bao}), the impact of RSD increases very fast at low bias (decreasing $\alpha$) in all FoMs, because RSD include an additional term such that the signal of the correlations does not drop when bias tend to zero (as happens when we do not have RSD, see B2 in Fig. \ref{fig:fom_redshift_bamp_real}). On the contrary, in $\fom$ the impact of BAO improves with more bias because shot-noise has less effect, while in $\fomg$ is the opposite and for $\alpha \geq 1$ it has a negative effect, although the effect is tiny. $\fomc$ (top panels) combines the effects from $\gamma$ and $\omega$ constraints, and shows that RSD has a bigger effect in the constraints than BAO.

\subsection{Fixing cosmological parameters} \label{subsubsec:xcosmo} 

In this subsection we investigate how fixing one cosmological parameter affects the forecast by breaking degeneracies. Fig. \ref{fig:fom_degen} shows the relative gain from fixing each parameter for B1xB2 (left panels) and B1 (right panels), for $\fomc$ and $\fomg$ on the rows, as a function of survey density. For $\fomc$ in B1xB2 there are strong gains (a factor 2$\sim$3) when fixing $\Omega_m$, $\Omega_{DE}$, $\Omega_b$ or $h$, which constrains the cosmic expansion history, while when fixing $\sigma_8$ and $n_s$ $\fomc$ only improves by a factor 1.2$\sim$1.3. Comparing to the single tracer, in $\fomc$ multiple tracers help breaking degeneracies for $\Omega_m$, $\Omega_b$, $\Omega_{DE}$ and $h$ parameters, specially at the noiseless limit where the relative gains lower from a factor $3.5\sim4$ in B1 to a factor 2 in B1xB2. For B1xB2 in $\fomg$ the gains are quite low ($<10$\%) when fixing any parameter, with $\sigma_8$ and $n_s$ the most relevant at low density, whereas at high density the gain from $\sigma_8$ drops and $\Omega_m$, $\Omega_{DE}$ and $h$ become more important. For B1 we find a similar level of degeneracy in $\fomg$.

\xfigure{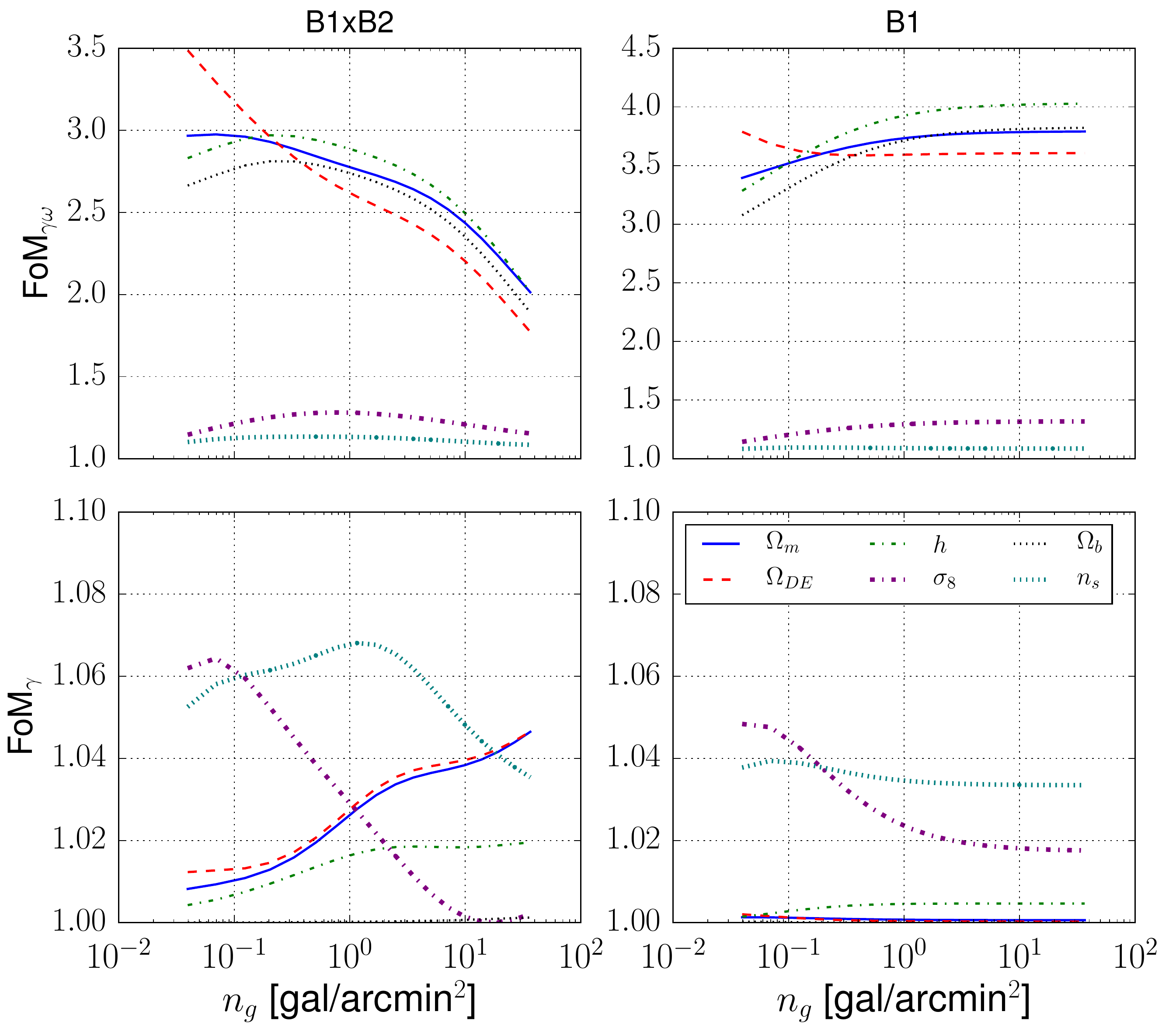}
{Relative gains when fixing a parameter for B1xB2 (left panels) and B1 (right panels), for $\fomc$ and $\fomg$ on the rows, as a function of survey density. All lines are normalized to the respective fiducial forecast values.}
{fig:fom_degen}

%% file: appendix_B_clean.tex
\section{Bias evolution}  \label{appB} 

In this section we study the impact that different bias evolution has on the constraints. For concreteness, we parametrize the bias evolution in redshift as

\begin{equation} \label{biaslope}
b_{B1}(z) = 1+ \kappa z,
\end{equation}

\medskip

\noindent where $\kappa=2$ corresponds to the fiducial forecast value, while we keep $\alpha=0.5$. Fig. \ref{fig:fom_zrange_biaslope} shows the FoMs for B1xB2 (left panels) and B1 (right panels) in the columns and $\fomc$ (top), $\fomg$ (center) and $\fom$ (bottom) for the rows, for different slope values, $\kappa\in[0.0, 0.5, 1.0, 1.5, 2.0]$ (red, green, blue, black, purple). $\fomdetf$ is very similar to $\fom$ and therefore not shown.

\xfigure{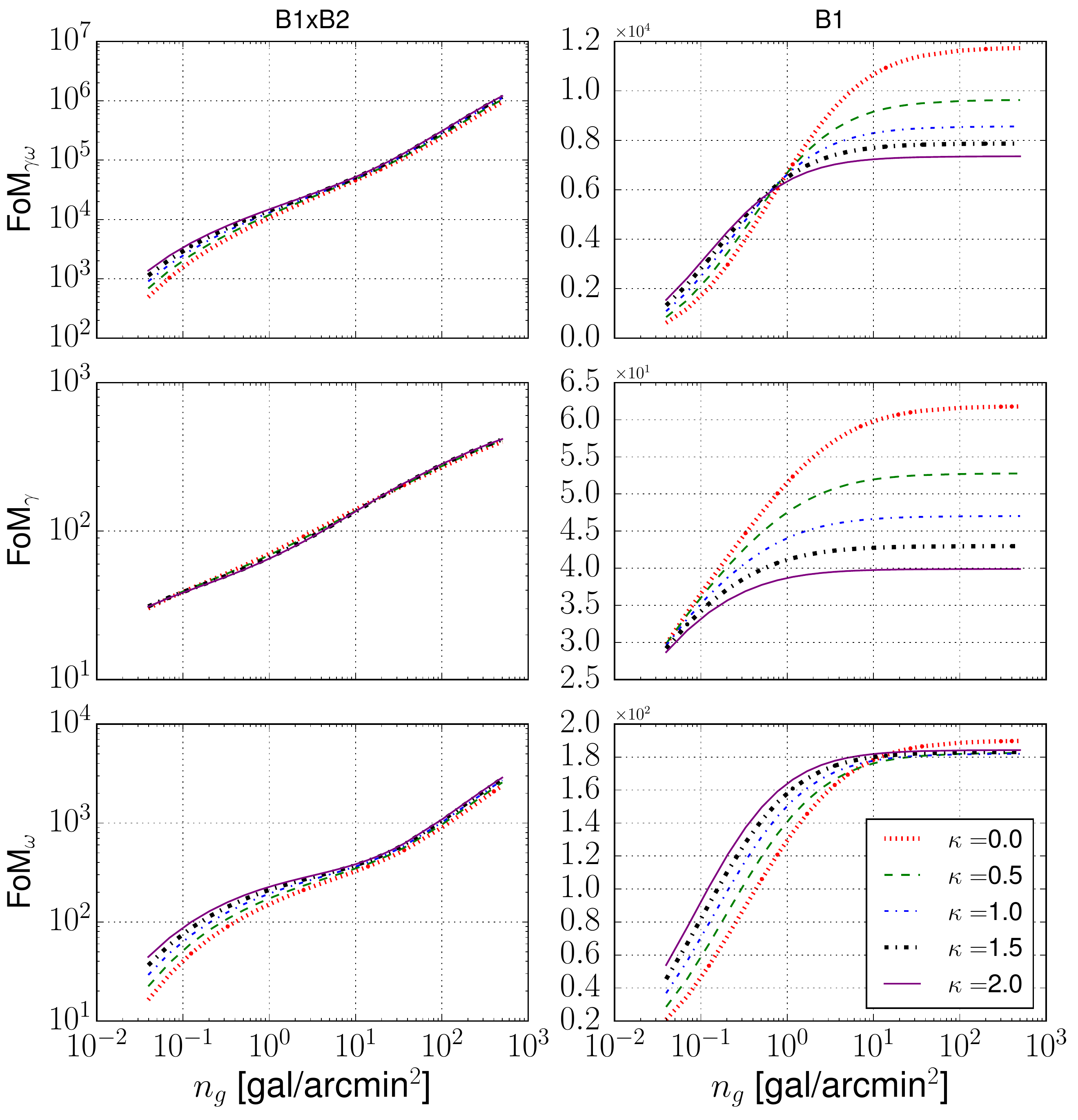}
{FoM dependence on the bias slope $\kappa$ ($b_{B1}(z) = 1+ \kappa z$), with $\alpha=0.5$ fixed. The panels on the left (right) shows the FoMs for B1xB2 (B1). On the rows we show $\fomc$ (top), $\fomg$ (center) and $\fom$ (bottom), with $\fomdetf$ very similar to $\fom$ and therefore not included. The lines correspond to $\kappa=0.0$ (red dotted), $\kappa=0.5$ (green dashed), $\kappa=1.0$ (blue dash-dot), $\kappa=1.5$ (black thicker dash-dot ), $\kappa=2.0$ (purple solid). }
{fig:fom_zrange_biaslope}

Looking at the multitracer panels (B1xB2) (Fig. \ref{fig:fom_zrange_biaslope}, left panels) it is clear that the constraints are not much affected by the bias evolution history. This shows that for the multi tracer it is much more relevant the relative bias amplitude between the populations rather than the bias evolution in redshift or the bias amplitude itself. Only at low densities we see some gains for higher bias, coming from the $\omega$ constraints, while the $\gamma$ constraints remain very similar for the different bias evolutions at all densities. In other words, the constraining power of B1xB2 does not rely on how biased are the samples themselves, but on the contrast between its subsamples. 

For the single tracer (B1) there is instead a clear dependence on the bias evolution in redshift (right panels). In $\fomg$ the constraints are much more affected by the different bias slopes at the noiseless limit, leading to better constraints for lower slope (and thus lower bias), as then RSD has more relative importance. For higher shot-noise the relative gains from RSD are lower as shot-noise dominates and the lines converge. The opposite happens in $\fom$, the dependence on bias slope is clear when shot-noise is big, while at the noiseless limit the constraints flatten and tend to the same value. As we have previosly remarked, dark energy constraints benefit from higher bias as this increases signal and reduces the relative impact of noise. This is observed clearly in this figures, as the FoMs benefit from higher bias at low densities, while with high densities they tend to the same value as the relative impact of noise is already small. $\fomc$ combines these effects and we see how all lines cross with each other, leading to better constraints for higher bias at low densities and viceversa at high densities. This results agree with Fig. \ref{fig:fom_redshift_bamp} and Fig. \ref{fig:foms_densityX}.  In addition, a complete unbiased tracer (i.e. $\kappa=0$) has the best constraints at the noiseless limit for all FoMs.